\documentclass{article}

\usepackage[english]{babel}

\usepackage[letterpaper,top=2.5cm,bottom=2.5cm,left=2cm,right=2cm,marginparwidth=1.75cm]{geometry}

\usepackage{tikz}
\usepackage{tikz-3dplot}
\usepackage{amsmath}
\usepackage{amsfonts}
\usepackage{amssymb}
\usepackage{graphicx}
\usepackage{siunitx}
\usepackage{braket}
\usepackage[normalem]{ulem}
\usepackage[colorlinks=true, allcolors=blue]{hyperref}
\usepackage{upgreek}
\usepackage{cleveref}
\usepackage{authblk}
\usepackage{dsfont} 
\usepackage{bm}
\usepackage[toc,page,title,titletoc]{appendix}
\usepackage[square,numbers,sort&compress]{natbib}
\usetikzlibrary{decorations.pathreplacing,decorations.pathmorphing,angles,quotes}

\begin{document}

\title{An optically-based universal gate set for hybrid continuous-variable and discrete-variable quantum computing on a trapped-ion processor}

\author[]{Edward C. Tortorici$^{*\dagger}$}
\author[]{Ethan C. McGarrigle$^*$}
\author[]{Brian K. McFarland}
\author[]{Wes L. Johnson}
\author[]{Daniel S. Lobser}
\author[]{Melissa C. Revelle}
\author[]{Brandon P. Ruzic}
\author[]{Susan M. Clark}
\author[]{Christopher G. Yale$^\ddagger$}
\affil[]{Sandia National Laboratories, Albuquerque, New Mexico, USA}

\newcommand{\ketbra}[2]{\ket{#1}\mkern-6mu\bra{#2}}
\newcommand{\proj}[1]{\ket{#1}\mkern-6mu\bra{#1}}
\newcommand{\identity}[0]{\hat{\mathds{1}}} 

\maketitle
\begin{NoHyper}
\def\thefootnote{*}\footnotetext{These authors contributed equally to this work}
\end{NoHyper}
\begin{NoHyper}
\def\thefootnote{$\dagger$}\footnotetext{Present address: IonQ Inc. 1685 38th St Unit 3, Boulder, CO 80301}
\end{NoHyper}
\begin{NoHyper}
\def\thefootnote{$\ddagger$}\footnotetext{Contact: cgyale@sandia.gov}
\end{NoHyper}

\begin{abstract}

    Here we present a universal gate set for hybrid continuous-variable and discrete-variable quantum computing for trapped-ion processors based on optical interactions. We implement these operations on the Quantum Scientific Computing Open User Testbed (QSCOUT). We generate these gates through variations of resonant and detuned motional-sideband interactions, and present state preparation and measurement methods. These operations include Jaynes-Cummings, Anti-Jaynes-Cummings, Conditional Displacement, Conditional Rotation, Conditional Beamsplitter, and Conditional (single-mode) Squeezing.

\end{abstract}

\section{Introduction}
    Fault-tolerant quantum computing promises to bring unparalleled capabilities for solving problems intractable for classical computers by leveraging physical quantum mechanical resources, such as trapped-ion, neutral atom, or superconducting qubits. However, fault-tolerant computing requires a significant qubit and hardware overhead to safekeep the quantum information in logical qubits and achieve computation error rates compatible with economic and scientific utility. Although the path to scaling fault-tolerant architectures to thousands of logical qubits is promising, no platform has yet reached the scale where utility is within reach. As a result, excitement remains surrounding alternative platforms and architectures that could achieve utility scale faster than traditional qubit-based approaches.
    
    One such alternative architecture takes a hybrid approach by storing quantum information in both discrete quantum variables \emph{and} continuous quantum variables. Discrete-variable (DV) quantum computing encodes information in a set of finite-level systems, such as two-level qubits or $d$-level qudits, and qubit-based quantum computing is currently the most common approach by industry efforts~\cite{bluvstein_logical_2024, aghaee_rad_scaling_2025, ransford_98-qubit_2026, acharya_quantum_2025, proctor_benchmarking_2026}. Continuous-variable (CV) quantum computing, on the other hand, encodes quantum information in the infinite-dimensional Hilbert space of bosonic modes~\cite{cv1, cv2}, which emerge fundamentally in the context of quantum harmonic oscillators such as optical modes, microwave cavities, and mechanical resonators. Hybrid CV-DV quantum computing combines advantages of CV and DV approaches. Continuous variables enable access to larger Hilbert spaces compared to qubit-only approaches, leading to more efficient resource usage. For example, when simulating Bose-Fermi mixture problems, fewer physical resources are required since the CV (DV) degrees of freedom naturally embed Bose (Fermi) statistics. Bose-Fermi couplings are found throughout nature, spanning fields ranging from high-energy physics~\cite{linke2026, yukawa,Bauer2023HEP,Bauer2023Fundamental,DiMeglio2024}, cold quantum gases \cite{foster_quantum_2026, desalvo_observation_2017}, and condensed matter physics~\cite{Devereese2015,Lanzara2001,Hebert2019,Ghosh2020}. Furthermore, hybrid CV-DV has the advantage over pure CV quantum computing because the hybrid qubit-motion interactions (discussed in this paper) enable the creation of non-linearities in the quantum harmonic oscillator~\cite{tan1,tan2,mcgarry,triggate}. 
    
    While bosonic resources can be engineered and coupled to qubit resources, such as microwave cavities coupled to superconducting qubits~\cite{putterman_hardware-efficient_2025, krastanov_universal_2015, eickbusch_fast_2022}, trapped ions have the advantage of having inherently built-in bosonic resources in the form of the ion chain's phonon modes that are accessible \emph{intrinsically} alongside qubit resources~\cite{sbc, bsb-readout, dynamics, hou_coherent_2024, chen_scalable_2023}. The ions' trapping environment and inter-ion Coulomb repulsions lead to a motional equilibrium described by normal modes of vibration. These shared phonon modes are routinely leveraged to drive high-fidelity qubit entangling gates via sideband couplings~\cite{ms1,ms2,sorensen_entanglement_2000,oi-ms}. Furthermore, coherent operations explicitly on the ions' motional modes, such as mode squeezing and displacement protocols, have grown in sophistication and maturity in recent years~\cite{sutherland2021, fluhmann-readout}. As a result, trapped ions have emerged as a particularly appealing platform for hybrid qubit-bosonic quantum information processing, and the large bosonic Hilbert space enables remarkably efficient encodings of quantum information per ion~\cite{fluhmann_encoding_2019}. 

    A successful CV-DV approach must achieve a universal physical and logical gate set, which has been outlined in recent efforts to bring CV-DV computing on more solid ground~\cite{yuan, araz_hybrid_2025}. In analogy to Clifford and non-Clifford gates in DV computing, CV computing requires Gaussian motional operations, such as mode displacements, squeezing, and rotation operations, as well as non-Gaussian interactions to achieve universality. Non-Gaussian operations such as Kerr-type interactions and tri-mode squeezing are particularly challenging to achieve in modern qubit-boson experiments due to coherence limitations or complex hardware requirements. However, hybrid CV-DV computing remarkably does not require explicit non-Gaussian interactions or non-Clifford gates, since compositions of operations on the joint qubit-boson Hilbert space, such as conditional displacements, can achieve the necessary nonlinearity for universal computation~\cite{kemper_hybrid_2025}.
    
    Here, we present an experimental implementation of a universal gate set for CV-DV quantum computing using the Quantum Scientific Computing Open User Testbed (QSCOUT)~\cite{qscout} using optical, laser-based interactions with trapped $^{171}$Yb ions. Our gate-set implementation consists of qubit-only operations and motional operations conditioned on the qubit degrees of freedom, such as conditional displacements, conditional squeezing, conditional rotations, and conditional beamsplitter operations. All qubit-motion interactions leverage first-order sideband transitions in the trapped-ion chain. In our implementation of the gate set, we use sideband cooling followed by motional sideband ladders to prepare desired Fock states natively. Furthermore, our implementation allows for native measurements of a motional mode's characteristic function and Fock state populations. In this paper, we present the fundamental qubit-motion operations and how these are programmed as gates on QSCOUT using Just Another Quantum Assembly Language, Jaqal$^\text{TM}$~\cite{jaqal}.

\section{Experimental System}
    QSCOUT is a trapped-ion quantum processor that utilizes a single, linear chain of $^{171}$Yb ions as a qubit register with all-to-all connectivity. The qubit states used are
\begin{equation}
\begin{split}
    \ket{{}^2\text{S}_{1/2}, F=0,m_F=0} &=\ket{g}\\
    \ket{{}^2\text{S}_{1/2}, F=1,m_F=0} &=\ket{e},
\end{split}
\end{equation}
    where $\ket{g}$ labels the lower energy state and $\ket{e}$ describes the higher energy state. The Pauli operators are defined such that
\begin{equation}
    \label{eq:z-convention}
    \sigma_z = \proj{e} - \proj{g}.
\end{equation}
    
    The eigenmodes of the ion chain's collective motion (motional modes or phonon modes) are utilized as quantum harmonic oscillators (qumodes). To access the qumodes, a Raman transition is driven with two counter-propagating beams, each orthogonal to the axial direction of the ion chain. On one side, there is a large beam that globally addresses the ion chain, and on the other side, there are tightly focused beams that individually address each ion~\cite{qscout}. Each beam can have up to four laser tones where each tone can be independently controlled with amplitude, phase, and frequency modulation. QSCOUT supports pulse-level control in Jaqal via the JaqalPaw interface~\cite{jaqalpaw}.

    An $N$ ion register has $N$ flavors of collective motion (see \Cref{fig:modes}), each manifesting three modes (one for each principal axis) for a total of $3N$ modes. Because the gate lasers point orthogonal to the axial direction of the ion chain, only the $2N$ radial modes are accessible to the optical-based gates. Collective motion encompasses the entire chain, but the coupling strengths of each ion to each mode are not uniform, and some ions have very weak to zero coupling to some modes.

\begin{figure}[htb]
    \centering
    \begin{tikzpicture}
        \node at (0,0) {\includegraphics{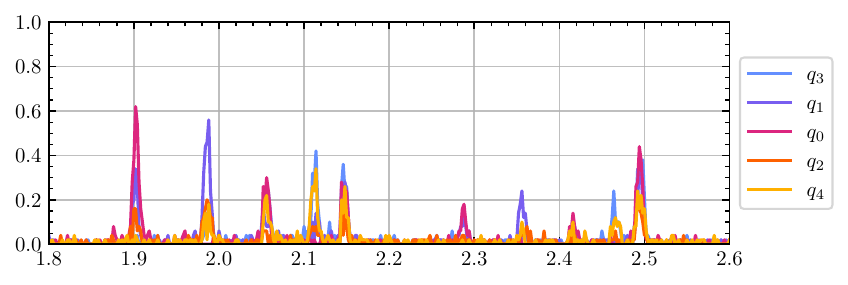}};
        \node[rotate=90] at (-7.3,0.2) {Detection Probability};
        \node at (0, -2.4) {Carrier Detuning Frequency $\omega_L-\omega_q$ (MHz)};
        \draw[decoration={brace,raise=5pt},decorate] (-5.3,2.2) --++ (4.3,0) node[midway, above, yshift=10,rounded corners=5pt, thick, fill=white] {$\mu=1$};
        \draw[decoration={brace,raise=5pt},decorate] (0.2,2.2) --++ (4.2,0) node[midway, above, yshift=10,rounded corners=5pt, thick, fill=white] {$\mu=0$};
        \draw (-5.1, .9) node[rounded corners=5pt, thick, fill=white] {$m=4$} ++
        (1.4, 0) node[rounded corners=5pt, thick, fill=white] {3} ++
        (1.05, -1) node[rounded corners=5pt, thick, fill=white] {2} ++
        (0.9, .3) node[rounded corners=5pt, thick, fill=white] {1} ++
        (.55, -.1) node[rounded corners=5pt, thick, fill=white] {0}
        (.6, -.5) node[rounded corners=5pt, thick, fill=white] {4} ++
        (1., .2) node[rounded corners=5pt, thick, fill=white] {3} ++
        (0.9, -.3) node[rounded corners=5pt, thick, fill=white] {2} ++
        (0.7, .3) node[rounded corners=5pt, thick, fill=white] {1} ++
        (.55, .7) node[rounded corners=5pt, thick, fill=white] {0};
    \end{tikzpicture}
    \caption{Motional sideband spectrum for 5 ions, each labelled $q_{k}$. The 5 flavors of radial modes each appear twice: one at a lower frequency (the lower manifold $\mu=1$), and one at a higher frequency (the higher manifold $\mu=0$). A qumode is then defined both by a manifold and a mode index. The axial frequency manifold does not appear because the lasers do not project onto these modes.}
    \label{fig:motional-sidebands}
\end{figure}

    Each flavor of radial mode can be indexed from 0 to $N-1$, as seen in \Cref{fig:modes}. In Jaqal, this index $m$ is referred to as \texttt{<mode>}, and these are counted from the highest frequency to the lowest frequency mode. The trap confinement is elliptical (see \Cref{fig:confinement}), such that one set of these $N$ radial modes is higher energy than the other. We refer to these as manifolds $\mu$ and are called out in Jaqal with \texttt{<manifold>}, where 0 corresponds to the higher and 1 to the lower frequency manifold. Each qumode then has a unique energy that can be denoted with a frequency $\omega_{\mu,m}$, where $\mu$ is the manifold index, and $m$ is the mode index.  The ellipticity of the confinement is selected such that $\omega_{0,i}>\omega_{1,j}$ for all $i$ and $j$. In other words, all modes of the upper manifold are higher energy than any mode in the lower manifold. An example of the motional sideband spectrum for a five-ion chain and how each qumode is identified is shown in \Cref{fig:motional-sidebands}.

\begin{figure}[t!]
    \centering
    \begin{tikzpicture}
        \tikzset{circ/.style={circle,draw=blue!50, fill=blue!10, very thick, minimum size=.1cm}}
        \tikzset{circ2/.style={circle,draw=orange!50, fill=orange!10, dashed, very thick, minimum size=.1cm}}
        \def\d{0.5}; \def\s{1.8};
        \draw[decoration={brace,raise=5pt},decorate] (-0.9,-.1) --++ (0,\d*5+.2) node[midway, xshift=-17,rotate=90] {qubit/ion index};
        \draw[dashed] (0,0) node[below, yshift=-9]{0} ++ (\s/6,0) node[circ2]{} --++ (0,\d) node[circ2]{} --++ (0,\d) node[circ2]{} --++ (0,\d) node[circ2]{} --++ (0,\d) node[circ2]{} --++ (0,\d) node[circ2]{};
        \draw (0,0) node[below, yshift=-9]{0} ++ (-\s/6,0) node[circ]{} node[xshift=-15]{4} --++ (0,\d) node[circ]{} node[xshift=-15]{2} --++ (0,\d) node[circ]{} node[xshift=-15]{0} --++ (0,\d) node[circ]{} node[xshift=-15]{1} --++ (0,\d) node[circ]{} node[xshift=-15]{3} --++ (0,\d) node[circ]{} node[xshift=-15]{5};
        \draw[latex-] (-.5,-.55) --++ (-.5,0) node[anchor=east] {mode index};

        \def\shift{1};
        \draw[dashed] (\s,5*\d) ++ (-\shift/2,0) node[circ2]{} --++ (\shift/6,-\d) node[circ2]{} --++ (\shift/6,-\d) node[circ2]{} --++ (\shift/6,-\d) node[circ2]{} --++ (\shift/6,-\d) node[circ2]{} --++ (\shift/6,-\d) node[circ2]{};
        \draw (\s,0) node[below, yshift=-9]{1} ++ (-\shift/2,0) node[circ]{} --++ (\shift/6,\d) node[circ]{} --++ (\shift/6,\d) node[circ]{} --++ (\shift/6,\d) node[circ]{} --++ (\shift/6,\d) node[circ]{} --++ (\shift/6,\d) node[circ]{};

        \def\shift{0.7};
        \draw[dashed] (2*\s,5*\d) ++ (-\shift/2+\shift/8,0) node[circ2]{} --++ (\shift/2,-\d) node[circ2]{} --++ (\shift/2,-\d) node[circ2]{} --++ (0,-\d) node[circ2]{} --++ (-\shift/2,-\d) node[circ2]{} --++ (-\shift/2,-\d) node[circ2]{};
        \draw (2*\s,0) node[below, yshift=-9]{2} ++ (\shift/2,0) node[circ]{} --++ (-\shift/2,\d) node[circ]{} --++ (-\shift/2,\d) node[circ]{} --++ (0,\d) node[circ]{} --++ (\shift/2,\d) node[circ]{} --++ (\shift/2,\d) node[circ]{};

        \def\shift{0.5};
        \draw[dashed] (3*\s,5*\d) ++ (-\shift/2,0) node[circ2]{} --++ (\shift,-\d) node[circ2]{} --++ (-\shift/3,-\d) node[circ2]{} --++ (-\shift/3,-\d) node[circ2]{} --++ (-\shift/3,-\d) node[circ2]{} --++ (\shift,-\d) node[circ2]{};
        \draw (3*\s,0) node[below, yshift=-9]{3} ++ (-\shift/2,0) node[circ]{} --++ (\shift,\d) node[circ]{} --++ (-\shift/3,\d) node[circ]{} --++ (-\shift/3,\d) node[circ]{} --++ (-\shift/3,\d) node[circ]{} --++ (\shift,\d) node[circ]{};
        
        \def\shift{0.5};
        \draw (4*\s,0) ++ (-\shift/2,0) node[circ2]{} --++ (\shift,\d) node[circ2]{} --++ (-\shift,\d) node[circ2]{} --++ (0,\d) node[circ2]{} --++ (\shift,\d) node[circ2]{} --++ (-\shift,\d) node[circ2]{};
        \draw (4*\s,0) node[below, yshift=-9]{4} ++ (\shift/2,0) node[circ]{} --++ (-\shift,\d) node[circ]{} --++ (\shift,\d) node[circ]{} --++ (0,\d) node[circ]{} --++ (-\shift,\d) node[circ]{} --++ (\shift,\d) node[circ]{};

        \def\shift{0.5};
        \draw (5*\s-\shift/6,5*\d) node[circ2]{} -- (5*\s+\shift/3,4*\d) node[circ2]{} -- (5*\s-\shift/2,3*\d) node[circ2]{} -- (5*\s+\shift/2,2*\d) node[circ2]{} -- (5*\s-\shift/3,\d) node[circ2]{} -- (5*\s+\shift/6,0) node[circ2]{};
        \draw (5*\s,0) node[below, yshift=-9]{5} (5*\s-\shift/6,0) node[circ]{} -- (5*\s+\shift/3,\d) node[circ]{} -- (5*\s-\shift/2,2*\d) node[circ]{} -- (5*\s+\shift/2,3*\d) node[circ]{} -- (5*\s-\shift/3,4*\d) node[circ]{} -- (5*\s+\shift/6,5*\d) node[circ]{};
    \end{tikzpicture}
    \caption{The flavors of radial modes for 6 ions labeled by mode index. Mode 0---the highest energy mode within a manifold---is the center of mass (COM) mode, and mode 1 is the the tilt mode. We refer to mode 2 as the drum mode, and the lowest energy mode ($N-1$) as the zig-zag mode (The remaining modes do not have strong naming conventions as they become more zig-zag-like in nature as the mode number increases). The qubits/ions are indexed starting from the center ion: they are labeled with increasing qubit number, alternating outward from the center, as seen on the left.}
    \label{fig:modes}
\end{figure}
\begin{figure}[t!]
    \centering
    \begin{tikzpicture}
        \node at (-1.9,1.6) {Radial Confinement};
        \draw[rotate=45] (0,0) ellipse (2cm and 1.1cm);
        \draw[rotate=45,stealth-stealth] (-2,0) node[anchor=north east] {$\omega_{1,j}$} -- (2,0);
        \draw[rotate=45,stealth-stealth] (0,-1.1) node[anchor=north west] {$\omega_{0,i}$} -- (0,1.1);
        \draw[thick,->] (-3,0) -- (-2,0) node[midway, below] {$\omega_\text{GB}$};
        \draw[thick,->] (3,0) -- (2,0) node[midway, below] {$\omega_{\text{IA},k}$};
    \end{tikzpicture}
    \caption{The radial confinement is elliptical, so that the two radial modes of the same mode index have different energies, such that $\omega_{0,i}>\omega_{1,j}$ for all $i$ and all $j$, where $i$ is the mode index for the upper manifold and $j$ is the mode index for the lower manifold. The Raman transition is driven with a globally addressing beam with frequency $\omega_\text{GB}$ and individually adressing beams with up to four frequencies $\omega_{\text{IA},k}$. The beat notes are then $\omega_{L,k}=\omega_{\text{IA},k}-\omega_\text{GB}$. The two beams driving Raman transitions are orthogonal to the axial direction, so they only project onto the radial modes.}
    \label{fig:confinement}
\end{figure}

\subsection{State Preparation and Measurement}

    On QSCOUT, each quantum circuit is prepared by optically pumping the ions to prepare them in the $\ket{g}$ state as described in reference~\cite{qscout}. This is followed by sideband cooling~\cite{sbc,dynamics} to prepare a thermal state very near the vacuum state $\ket{0}$. We typically measure an average occupation of about 0.1 quanta after sideband cooling in non-COM modes using sideband thermometry~\cite{sbt}. Sideband cooling is performed by selecting which modes to cool on which ions, and then applying an approximate $\pi$-pulse on the red-sideband to drive of the associated modes on each of the qubits to drive them from $\ket{g}$ to $\ket{e}$ while removing a quanta from the motional state (see \Cref{sec:jc}). This is followed by an optical pumping step to return the qubit to $\ket{g}$. Repeating this process many times, will bring the motional states close to $\ket{0}$ preparing an initial state of $\ket{g}^{\otimes N}\otimes\ket{0}^{\otimes 2N}$, where there are $2N$ accessible radial modes in the $N$-ion chain.

    Qubit measurement is performed by scattering light from the $\ket{e}$ state and described in detail in Section V of reference~\cite{qscout}. Here, we report $P$ as the probability of qubit state $\ket{\psi}$ being in the bright state $\ket{e}$: $P=|\braket{e|\psi}|^2$. Measurement of a motional state is performed by entangling it to the qubit through one of the hybrid qubit-motion gates described in this paper and then measuring the qubit state~\cite{fock-prep,bsb-readout,fluhmann-readout,wigner-readout,wigner-readout2,state-reconstruction,blatt-tomography,Winkelmann_2022}. We have demonstrated two motional state readout methods on QSCOUT: (1) where an anti-Jaynes-Cummings operation is performed on the ion such that qubit population as a function of the interaction strength can be used to determine the Fock occupancies (described in \Cref{sec:fock-readout})~\cite{fock-prep,bsb-readout} and (2) using a spin-dependent force in a way that the expectation value for the $\sigma_z$ operator on the qubit relates directly to the expectation value for the displacement operator on the motional state (described in \Cref{sec:motional-tomography})~\cite{fluhmann-readout}.
    
\begin{figure}[htb]
    \centering
    \begin{tikzpicture}
        \node[anchor=south west, inner sep=0] (figa) at (0,0)
            {\includegraphics[]{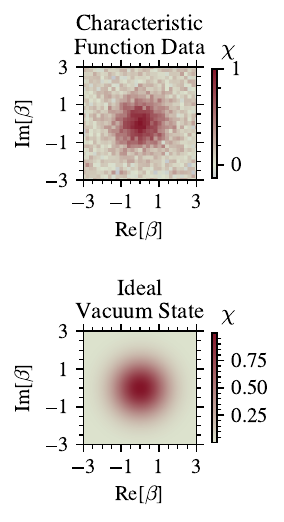}};
        \node[anchor=south west, inner sep=0] (figb) at (0.32\textwidth,0)
            {\includegraphics{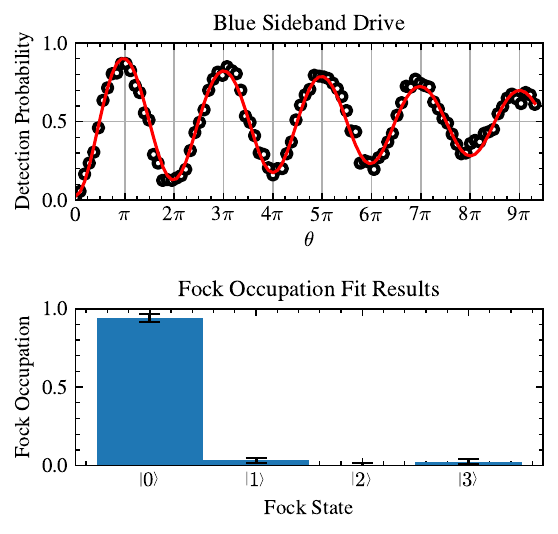}};
        \draw (-.5, 8) node {(a)}
            (-.5, 4) node {(b)}
            (5.25, 8) node {(c)}
            (5.25, 4) node {(d)};
    \end{tikzpicture}
    \caption{Characterization after sideband cooling.
    (a) The measured characteristic function $\chi(\beta)$ after sideband cooling is performed, and (b) the ideal characteristic function for a vacuum state as described by \Cref{eq:chi-vac}. The fidelity calculated from this data is $0.96$. The process for how this type of data is collected and how the fidelity is calculated is discussed in \Cref{sec:motional-tomography}. (c) Blue sideband readout after sideband cooling. This data is fitted as described in \Cref{sec:jc} to extract Fock occupations, depicted in d). The fit yields an occupation in the $\ket{0}$ state of $0.941 \pm 0.028$, leaving some residual population in higher Fock states.
    }
    \label{fig:vacuum-characterization}
\end{figure}

    The preparation of the vacuum state through sideband cooling can be characterized through both motional-state measurement methods. The characteristic function for the vacuum state is
\begin{equation}
    \label{eq:chi-vac}
    \chi_\text{vacuum}(\beta) = \bra{0}\mathcal{D}(\beta)\ket{0} = e^{-|\beta|^2/2},
\end{equation}
    where $\beta$ is a complex-valued parameter representing points in phase space with $\mathfrak{Re}[\beta]$ representing position and $\mathfrak{Im}[\beta]$ representing momentum of the oscillator. \Cref{eq:chi-vac} is plotted in \Cref{fig:vacuum-characterization} alongside a measurement of the motional state after circuit preparation.

    As described in \Cref{sec:fock-readout}, the effective Rabi rate of the blue sideband drive on an initial state of $\ket{g,0}$ is $\Omega\eta_{\mu,m,k}$, where $\eta_{\mu,m,k}$ is the Lamb-Dicke factor for ion $k$'s participation in qumode $\mu,m$. For single-qumode-single-qubit process only one Lamb-Dicke factor is relevant, so we refer to $\eta_{\mu,m,k}\equiv\eta$ unless more than one is needed to be specified. The blue sideband Rabi oscillations, as seen in \Cref{fig:vacuum-characterization}(c), can be fit to characterize how much residual population is in higher Fock states. Motional states live in an infinite-dimensional Hilbert space, but for practical purposes, depending on the context, it is often necessary to work in a truncated basis. For example, when working in the Fock basis of phonon-occupation number states $\{\ket{n}, n \geq 0\}$, it becomes increasingly difficult to distinguish Fock states as $n$ increases, and truncating around $\ket{n=10}$ is recommended given current readout limitations \cite{yukawa}.

\subsection{The Light-Matter Interaction}

    We drive the qubit with Raman transitions from a \SI{355}{\nano\meter} pulsed laser, as described in Section IV of~\cite{qscout}, to create a beat frequency $\omega_L$ via virtual states in the $P_{1/2}$ and $P_{3/2}$ electronic manifolds of $^{171}$Yb. We consider an effective Hamiltonian of the qubit states in the dipole approximation, where the relative phase between the beams sets the phase $\phi$ of the effective coupling. In the Lamb-Dicke approximation and only considering coupling of an ion with a single qumode with manifold index $\mu$, mode index $m$, and frequency $\omega_{\mu,m}$, the interaction Hamiltonian~\cite{dynamics} for the ion-mode coupling can be written
\begin{equation}
    \label{eq:int-h}
    \hat{H}_\text{int} = \frac{\hbar}{2}\Omega\hat\sigma_+\left[
        1+i\eta\left(
            \hat{a}e^{-i\omega_{\mu,m}t} + \hat{a}^\dagger e^{i\omega_{\mu,m}t}
        \right)
    \right]e^{-i(\phi+(\omega_L-\omega_q) t)} + \text{h.c.}
\end{equation}
    where $\Omega$ is the two-photon Rabi frequency, $\eta$ is the Lamb-Dicke factor describing the ion's participation in the normal motion for mode $m$ from manifold $\mu$, $\hat{\sigma}_\pm = (\hat{\sigma}_x\pm i\hat{\sigma}_y)/2$ are the spin raising and lowering operators that act on the two-level system, and $a$ and $a^\dagger$ are the bosonic annihilation and creation operators for the mode of interest. This Hamiltonian is representative of cases where the beat frequency $\omega_L$ is tuned near a resonance of interest, such that the dynamics of spectator modes and other atomic levels may be omitted from the essential model but studied further during error analysis.

    The interaction Hamiltonian has three resonances: the carrier qubit transition $\omega_L \to \omega_q$ and the first motional sideband transitions $\omega_L \to \omega_q\pm\omega_{\mu,m}$. We consider \Cref{eq:int-h} in the limit where the beat frequency is tuned near each resonance. When $\omega_L$ is within some small detuning $\delta$ from any of these resonances and the overall coupling strength is much smaller than $\omega_{q}$, the rotating-wave approximation can be applied to neglect far off-resonant processes in \Cref{eq:int-h}. Considering each resonance as separate cases, we define the governing Hamiltonian for each: 
\begin{align}
    \label{eq:car-h}
    \hat{H}_\text{C} &= \frac{\hbar}{2}\Omega\left(\hat{\sigma}_+e^{-i(\phi+\delta t)}+\hat{\sigma}_-e^{i(\phi+\delta t)}\right)\\
    \label{eq:rsb-h}
    \hat{H}_\text{RSB} &= \frac{\hbar}{2}\Omega\eta\left(\hat{\sigma}_+\hat{a}e^{-i(\phi-\frac{\pi}{2}+\delta t)}+\hat{\sigma}_-\hat{a}^\dagger e^{i(\phi-\frac{\pi}{2}+\delta t)}\right)\\
    \label{eq:bsb-h}
    \hat{H}_\text{BSB} &= \frac{\hbar}{2}\Omega\eta\left(\hat{\sigma}_+\hat{a}^\dagger e^{-i(\phi-\frac{\pi}{2}+\delta t)}+\hat{\sigma}_-\hat{a} e^{i(\phi-\frac{\pi}{2}+\delta t)}\right),
\end{align}
    where \Cref{eq:car-h} describes the carrier drive, \Cref{eq:rsb-h} the red sideband drive, and \Cref{eq:bsb-h} the blue sideband drive. Note that the $\pi/2$ appears due to the factor of $i$ in front of the Lamb-Dicke parameter in \Cref{eq:int-h}. 

    Single qubit rotation operations are performed using a resonant carrier drive ($\delta \to 0$) for a time $t$:
\begin{equation}
\label{eq:sqr}
    \hat{R}(\theta,\phi)=e^{-i H_\text{C}t/\hbar}=\exp\left[
        -i\frac{\theta}{2}\hat\sigma_\phi\right],
\end{equation}
    where the rotation amplitude $\theta=\Omega t$ and $\hat\sigma_\phi=\hat\sigma_+e^{-i\phi}+\hat\sigma_-e^{i\phi}=\hat\sigma_x\cos\phi+\hat\sigma_y\sin\phi$ defines the rotation axis about the equator of the Bloch sphere. The eigenstates of $\hat{R}(\theta,\phi)$ are
\begin{equation}
    \ket{\pm;\phi} = \frac{1}{\sqrt{2}}\left(\ket{e}\pm e^{i\phi}\ket{g}\right).
\end{equation}

    All the hybrid-motion gates described in this paper are based on varying combinations of blue and red sideband drives. Since effective Rabi rates $\Omega_\text{eff}=\Omega\eta$ depend on the Lamb-Dicke parameter for the target ion-mode pair, $\eta$ must be determined during calibration. QSCOUT offers two options effective Rabi rate determination: (1) calculating $\eta$ numerically by an equilibrium analysis of the secular frequencies experienced by the ion chain~\cite{noauthor_sandialabsionsim_nodate}, and (2) direct calibration of each $\eta$ for all relevant ion-mode pairs by driving and detection probabilities as a function of sideband drive times and fitting the Rabi oscillations for $\eta$. However, direct calibration of the Lamb-Dicke factors can be cumbersome for qubit-motion circuits using many ions and many modes.

\section{Jaynes-Cummings and Anti-Jaynes-Cummings Gates}\label{sec:jc}
\begin{figure} 
\includegraphics{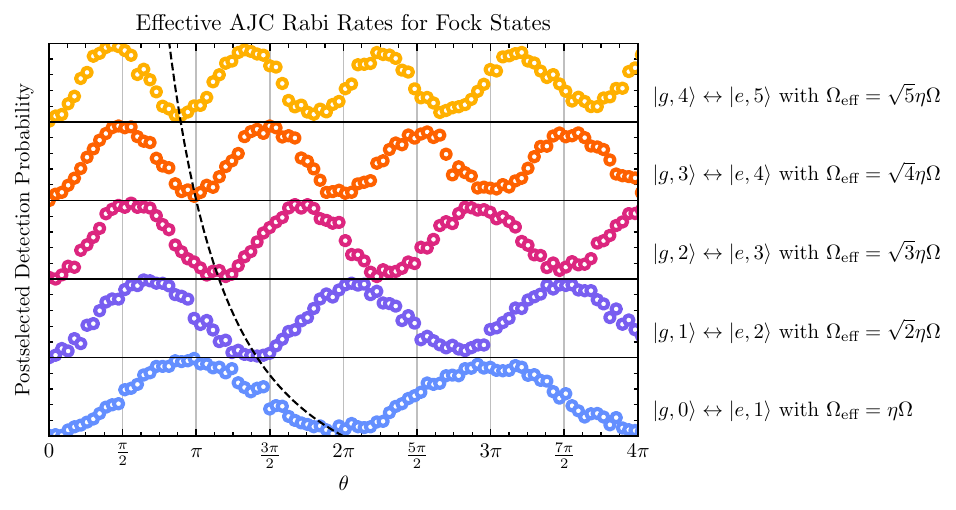}
    \caption{The effective Rabi rate $\Omega_{\rm eff}$ for the AJC gate scales with $\sqrt{n+1}$ and the Lamb-Dicke factor $\eta$ for target mode $m$ of manifold $\mu$. Here Fock states from $\ket{0}$ to $\ket{4}$ are driven at varying gate angle to demonstrate this dependence. The dashed line plots $(2\pi/\Omega t)^2-1$ to show where the first full Rabi cycle should occur for each Fock state.}
    \label{fig:eff-rabi-fock}
\end{figure}

    The dynamics of the pulses tuned near a red or blue sideband process leads to realization of the Jaynes-Cummings (JC) and Anti-Jaynes-Cummings (AJC) models. JC and AJC operations describe interactions between a two-level system and a bosonic mode with unitary operations
\begin{align}
    \hat{U}_\text{JC}(\theta, \phi) &= \exp\left[-i\frac{\theta}{2}( \hat{\sigma}_+\hat{a} e^{-i\phi}+\hat{\sigma}_-\hat{a}^\dagger e^{i\phi})\right] \\
    \hat{U}_\text{AJC}(\theta, \phi) &= \exp\left[-i\frac{\theta}{2}(\hat{\sigma}_+\hat{a}^\dagger e^{-i\phi}+ \hat{\sigma}_-\hat{a} e^{i\phi})\right].
\end{align}
    We generate these interactions using a resonant red sideband (\Cref{eq:rsb-h} with $\delta\rightarrow 0$) for the JC gate, and a resonant blue sideband (\Cref{eq:bsb-h} with $\delta\rightarrow 0$) for the AJC gate:
\begin{align}
    \hat{U}_\text{RSB}\left(t,\;\phi_\text{SB}=\phi+\frac{\pi}{2}\right) 
    &= \exp\left[-i \frac{\hat{H}_\text{RSB}}{\hbar} t\right] 
    = \exp\left[-i\frac{\eta \Omega t}{2}(\hat{\sigma}_+\hat{a} e^{i\phi}+\hat{\sigma}_-\hat{a}^\dagger e^{-i\phi})\right] \\
    \hat{U}_\text{BSB}\left(t,\;\phi_\text{SB}=\phi+\frac{\pi}{2}\right) 
    &= \exp\left[-i \frac{\hat{H}_\text{\text{BSB}}}{\hbar} t\right] 
    = \exp\left[-i\frac{\eta \Omega t}{2}(\hat{\sigma}_+\hat{a}^\dagger e^{i\phi}+\hat{\sigma}_-\hat{a} e^{-i\phi})\right],
\end{align}
    where $\phi_\text{SB}$ is the phase applied to the sideband operation, and $\theta=\eta \Omega t$ is the gate angle. Note that the input phase to the sideband drive $\phi_\text{SB}$ is shifted by $\pi/2$ to get a target gate phase of $\phi$. These unitary evolutions can be rewritten
\begin{align}
    \hat{U}_\text{RSB} &= \sum_{k=0}^\infty \frac{(-1)^k}{(2k)!}\left(\frac{\theta}{2}\right)^{2k}\left(
        \proj{g}\hat{N}^k+\proj{e}(\hat{N}+1)^k
    \right) - \frac{i(-1)^k}{(2k+1)!}\left(\frac{\theta}{2}\right)^{2k+1}\left(
        \hat{\sigma}_+(\hat{N}+1)^k\hat{a}+\hat{\sigma}_-\hat{N}^k \hat{a}^\dagger
    \right)\\
    \hat{U}_\text{BSB} &= \sum_{k=0}^\infty \frac{(-1)^k}{(2k)!}\left(\frac{\theta}{2}\right)^{2k}\left(
        \proj{g}(\hat{N}+1)^k+\proj{e}\hat{N}^k
    \right) - \frac{i(-1)^k}{(2k+1)!}\left(\frac{\theta}{2}\right)^{2k+1}\left(
        \hat{\sigma}_+\hat{N}^k \hat{a}^\dagger+\hat{\sigma}_- (\hat{N}+1)^k\hat{a}
    \right),
\end{align}
    and these act on states where the qubit is $\ket{g}$ and the motion is a Fock state in the following way
\begin{align}
    \hat{U}_\text{RSB}\left(t,\; \phi+\frac{\pi}{2}\right) \ket{g, n} &= \cos\left(\sqrt{n}\;\frac{\eta \Omega t}{2}\right) \ket{g, n} + e^{i(\phi-\pi/2)} \sin\left(\sqrt{n}\;\frac{ \eta \Omega t}{2}\right)\ket{e,n-1}\\
    \hat{U}_\text{BSB}\left(t,\; \phi+\frac{\pi}{2}\right) \ket{g,n-1} &= \cos\left(\sqrt{n}\;\frac{\eta \Omega t}{2}\right) \ket{g,n-1} + e^{i(\phi-\pi/2)} \sin\left(\sqrt{n}\;\frac{\eta \Omega t}{2}\right)\ket{e,n},
\end{align}
    where it can be seen that the red sideband couples $\ket{g,n}$ to $\ket{e,n-1}$ and the blue sideband couples $\ket{g,n-1}$ to $\ket{e, n}$, both with an effective Rabi rate that scales with $\sqrt{n}$ (see \Cref{fig:eff-rabi-fock}), which shows that the amount of rotation actually performed on the qubit will depend on the motional state.
    
    The implementation on QSCOUT is
\begin{align}
    \label{eq:JC}
    \hat{U}_\text{JC}(\theta, \phi) &= \hat{U}_\text{RSB}\left(t=\frac{t_\pi\theta}{\eta\pi},\phi_\text{SB}=\phi+\frac{\pi}{2}\right) \\
    \label{eq:AJC}
    \hat{U}_\text{AJC}(\theta, \phi) &= \hat{U}_\text{BSB}\left(t=\frac{t_\pi\theta}{\eta\pi},\phi_\text{SB}=\phi+\frac{\pi}{2}\right),
\end{align}
    where
    $t_\pi = \pi/\Omega$
    is a calibrated $\pi$-time for the carrier transition. 
    
\subsection{Anti-Jaynes-Cummings Operation for Measuring Fock State Occupation}\label{sec:fock-readout}

    Since the JC and AJC gates have effective Rabi rates that scale as $\sqrt{n}$, they can be leveraged to extract Fock occupancies of a motional state. Since the JC operation cannot drive a transition from $\ket{g,0}$, we use AJC for this task, as it will drive transitions from $\ket{g,n}$ for all $n$.

    An arbitrary motional state can be described in the Fock basis:
\begin{equation}
\label{eq:fock-basis}
    \ket{\psi_m} = \sum_{n=0}^\infty c_n \ket{n},
\end{equation}
    where $c_n$ are the complex-valued Fock amplitudes, and we can define the Fock occupation probabilities as the absolute squares of the Fock amplitudes $p_n = |c_n|^2$ which obey normalization. As a result,the probability of detecting $\ket{e}$ after driving the AJC gate for time $t$ on an arbitrary motional state is 
\begin{equation}
\label{eq:bsb_fit}
\begin{split}
    P(\theta) &= \frac{1}{2}\left[1 - e^{- \Gamma \theta} \sum_{n=0}^\infty p_n \cos\left(\sqrt{n+1}\ \theta\right)\right],
    \end{split} 
\end{equation}
    where $\Gamma = 1/\eta\Omega T_{2}$ is a general damping factor that captures potentially several dissipation mechanisms with differing character~\cite{dynamics}. For a single dominant Markovian dissipation mechanism, \Cref{eq:bsb_fit} can be recovered exactly via Lindblad master equation dynamics when $\eta \Omega \gg \Gamma$. By averaging the results of many shots at varying gate angles, $P(\theta)$ data can be generated and fit for $p_n$ for $n$ up until a reasonable cutoff.
    
\subsubsection{Example of AJC Measurement}
\begin{figure} 
\includegraphics{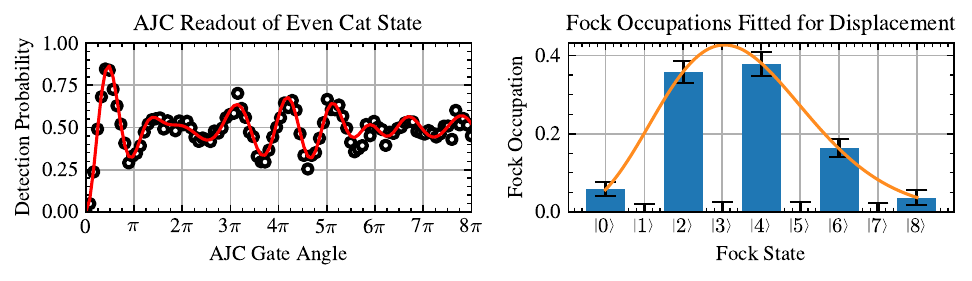}
    \caption{Example Fock state occupancy measurement. (left) The resultant beating oscillations from Fock occupation measurement via the AJC gate. The oscillations are fit to \Cref{eq:bsb_fit} to extract the Fock occupancies $p_n$ presented on the (right).}
    \label{fig:ajc_readout}
\end{figure}

    As an example, we prepare an even cat state and measure by scanning the $\hat{U}_\text{AJC}(\theta,\phi)$ gate over $\theta$. The even cat state is
\begin{equation}
    \ket{\psi}_\text{cat} = \ket{\alpha} + \ket{-\alpha} = 2e^{-|\alpha|^2/2} \sum_{k=0}^\infty \frac{\alpha^{2k}}{\sqrt{(2k)!}}\ket{2k},
\end{equation}
    where only even Fock states are populated. The distribution of even Fock populations follows a Poisson distribution. The dataset of the AJC scan for the even cat state is shown \Cref{fig:ajc_readout}. Fitting this curve, using \Cref{eq:bsb_fit} yields Fock occupancies, shown on the right of the figure. The fit puts virtually no population in the odd Fock states, and the orange curve shows how the even populations fit to a Poisson distribution. Here, we perform the fit with variable Fock cutoff, from $\ket{0}$ to $\ket{15}$ and pick the result with lowest cost per degree of freedom, where the cost per degree of freedom is
\begin{equation}
    F = \frac{1}{K-p}\sum_{k=1}^K(P_k-P(\theta_k))^2,
\end{equation}
    where $P_k$ are data points at $\theta_k$, $P(\theta)$ is described by \Cref{eq:bsb_fit} with the sum over $n$ going up to some Fock cutoff, $K$ is the number of data points, and $p$ is the number of free parameters in the fit.

\subsection{Fock State Preparation}\label{sec:fock-prep}
    To prepare a Fock state, after sideband cooling, alternated AJC and JC pulses are used to sequentially invert the qubit state while increasing the Fock state with each pulse~\cite{bsb-readout}. However, since the effective Rabi rate increases as the Fock state increases, each successive pulse is shortened to accommodate this. A so-called prep qubit is chosen to perform these operations on.

    Sideband cooling prepares $\ket{g, 0}$ for the relevant qubit and qumode, and each gate inverts the qubit state while increasing the Fock number by one, so possible states in the Fock state preparation procedure are either $\ket{g, 2k}$ or $\ket{e, 2k+1}$ with $k \in \mathbb{Z}^{+}$. During this process, the AJC gate is applied to $\ket{g, 2k}$ to increase even Fock numbers and excite the qubit:
\begin{equation}
    \hat{U}_\text{AJC}\left(\theta=\frac{\pi}{\sqrt{n+1}},\;\phi=0\right)\ket{g, n} = \ket{e, n+1},
\end{equation}
    while the Jaynes-Cummings gate is applied to $\ket{e, 2k+1}$ to increase odd Fock numbers and relax the qubit:
\begin{equation}
    \hat{U}_\text{JC}\left(\theta=\frac{\pi}{\sqrt{n+1}},\;\phi=0\right)\ket{e,n}=\ket{g,n+1}.
\end{equation}

    Preparing an even Fock state applies the following operations ending in a JC gate:
\begin{equation}
    \text{even }n:\ \ \hat{U}_\text{JC}\left(\frac{\pi}{\sqrt{n}}, 0\right)...\hat{U}_\text{AJC}\left(\frac{\pi}{\sqrt{3}},0\right)\hat{U}_\text{JC}\left(\frac{\pi}{\sqrt{2}},0\right)\hat{U}_\text{AJC}(\pi,0)\ket{g, 0} = \ket{g, n},
\end{equation}
    and preparing an odd Fock state ends with an AJC:
\begin{equation}
    \text{odd }n:\ \ \hat{U}_\text{AJC}\left(\frac{\pi}{\sqrt{n}}, 0\right)...\hat{U}_\text{AJC}\left(\frac{\pi}{\sqrt{3}},0\right)\hat{U}_\text{JC}\left(\frac{\pi}{\sqrt{2}},0\right)\hat{U}_\text{AJC}(\pi,0)\ket{g, 0} = \ket{e, n}.
\end{equation}
    Therefore, readout of the qubit state of the prep qubit can be used to confirm whether the motional state was prepared in an even or odd Fock state and this information can be used for postselection.

\section{Conditional Displacement Gate}\label{sec:cd}

    A conditional displacement (CD) operation is defined as
\begin{equation}
    \hat{U}_\text{CD}(\alpha, \mathbf{\hat{n}}) = \mathcal{\hat{D}}(\bm{\Vec{\sigma}}\cdot\mathbf{\hat{n}}\alpha) = \exp\left[\bm{\Vec{\sigma}}\cdot\mathbf{\hat{n}}(\alpha \hat{a}^\dagger-\alpha^* \hat{a})\right],
\end{equation}
    where $\alpha$ is the displacement amplitude, $\mathcal{\hat{D}}$ is the displacement operator, $\bm{\Vec{\sigma}}=(\sigma_x\mathbf{\hat{x}}+\sigma_y\mathbf{\hat{y}}+\sigma_z\mathbf{\hat{z}})$, and $\mathbf{\hat{n}}$ defines a unit vector direction on the Bloch sphere that sets the basis in which the displacement is conditioned. The conditional displacement can be reformulated
\begin{equation}
    \hat{U}_\text{CD}(\alpha, \mathbf{\hat{n}}) = \proj{+,\mathbf{\hat{n}}}\mathcal{\hat{D}}(\alpha) + \proj{-,\mathbf{\hat{n}}}\mathcal{\hat{D}}(-\alpha), \label{eq: CD_spin_expansion}
\end{equation}
    where $\ket{\pm,\mathbf{\hat{n}}}$ are the eigenkets of $\bm{\Vec{\sigma}}\cdot\mathbf{\hat{n}}$ with eigenvalues $\pm1$. In this form, it can be seen that the CD gate entangles the $+1$ eigenket with a displacement of $\alpha$ and the $-1$ eigenket with a displacement of $-\alpha$.

    To perform a conditional displacement gate, we apply a spin-dependent force (SDF)~\cite{patty}, where both a red and blue sideband are driven at the same time with equal and opposite detunings $\delta$ from the sidebands (such as being blue-detuned from the red sideband and red-detuned from the blue sideband) and respective phases $\phi_{r}$ (red) and $\phi_{b}$ (blue). This requires applying two tones to the individual addressing beam such that there are two relevant beat frequencies $\omega_{L,r}=\omega_q-\omega_{\mu,m}+\delta$ and $\omega_{L,b}=\omega_q+\omega_{\mu,m}-\delta$. This interaction has the following Hamiltonian:
\begin{equation}
\begin{split}
    \hat{H}_\text{SDF}(\phi_{q}, \phi_{m}) &= \hat{H}_\text{RSB}(\phi_r,\;\delta) + \hat{H}_\text{BSB}(\phi_b,\;-\delta)\\
    &= \frac{\hbar\eta \Omega}{2}\hat\sigma_{\phi_q}\left(
        \hat{a}^\dagger e^{i(\phi_m+\frac{\pi}{2}+\delta t)}+\hat{a}e^{-i(\phi_m+\frac{\pi}{2}+\delta t)}
    \right), \label{eq: SDF}
\end{split}
\end{equation}
    where $\delta$ is a symmetric detuning from the sidebands, $\phi_m$ is a phase related to the motion, and $\phi_q$ is a phase related to the qubit. These phases are related to the RSB and BSB phases by
\begin{align}
    \phi_q &= \frac{\phi_r+\phi_b+\pi}{2}\\
    \phi_m &= \frac{\phi_r-\phi_b+\pi}{2}.
\end{align}

    The unitary evolution of the SDF, when $\delta=0$, is then
\begin{equation}
\begin{split}
    \hat{U}_\text{SDF} &= \exp\left[-i\frac{\eta \Omega  t}{2}\hat{\sigma}_{\phi_q}\left(
        \hat{a}^\dagger e^{i(\phi_m+\frac{\pi}{2})} + \hat{a} e^{-i(\phi_m+\frac{\pi}{2})}
    \right)\right]\\
    &= \exp\left[\frac{\eta \Omega t}{2}\hat{\sigma}_{\phi_q}\left(
        \hat{a}^\dagger e^{i\phi_m} - \hat{a} e^{-i\phi_m}
    \right)\right].
\end{split}
\end{equation}
    By setting $\alpha=\eta \Omega t e^{i\phi_m}/2$, this becomes
\begin{equation}
\begin{split}
    \hat{U}_\text{SDF}(t_\alpha,\phi_b,\phi_r) 
    &= \exp\left[\hat{\sigma}_{\phi_q}\left(
        \alpha\hat{a}^\dagger - \alpha^* \hat{a}
    \right)\right],\\
    &= \proj{+;\phi_q}\hat{\mathcal{D}}(\alpha)+\proj{-;\phi_q}\hat{\mathcal{D}}(-\alpha),
\end{split}
\end{equation}
    with
\begin{align}
    t_\alpha &= \frac{2|\alpha|}{\eta \Omega}\\
    \phi_b &= \phi_q-\phi_m\\
    \phi_r &= \phi_q+\phi_m-\pi,
\end{align}
    which is a conditional displacement limited to being conditioned on the equator of the Bloch sphere. This can be generalized by wrapping the gate with single qubit rotations, where the eigenstates of an arbitrary $\bm{\Vec{\sigma}}\cdot\mathbf{\hat{n}}$ can be related to the eigenstates of $\sigma_{\phi_{q}}$ by
\begin{equation}
    \ket{\pm;\mathbf{\hat{n}}} = \hat{R}^\dagger\left(\frac{\pi}{2}-\theta_q,\;\frac{\pi}{2}+\phi_q\right)\ket{\pm;\phi_q},
\end{equation}
    where $\theta_q$ and $\phi_q$ are defined such that $\hat{\mathbf{n}}=(\sin\theta_q\cos\phi_q\;\mathbf{\hat{x}}+\sin\theta_q\sin\phi_q\;\mathbf{\hat{y}}+\cos\theta_q\;\mathbf{\hat{z}})$ so the generalized CD gate is
\begin{equation}
    \hat{U}_\text{CD}(\alpha, \mathbf{\hat{n}}) = \hat{R}^\dagger\left(\frac{\pi}{2}-\theta_q,\;\frac{\pi}{2}+\phi_q\right)
    \hat{U}_\text{SDF}(t,\phi_b,\phi_r) 
    \hat{R}\left(\frac{\pi}{2}-\theta_q,\;\frac{\pi}{2}+\phi_q\right).
\end{equation}

    On QSCOUT, there are also CD gates conditioned on each of the Paulis available:
\begin{align}
    \hat{U}_{x\text{CD}}(\alpha) &= \hat{\mathcal{D}}(\hat\sigma_x\alpha) = \hat{U}_\text{SDF}(t_\alpha,\; \phi_b=-\phi_m,\; \phi_r=\phi_m+\pi)\\
    \hat{U}_{y\text{CD}}(\alpha) &= \hat{\mathcal{D}}(\hat\sigma_y\alpha) = \hat{U}_\text{SDF}\left(t_\alpha,\;\phi_b=\frac{\pi}{2}+\phi_m,\;\phi_r=\frac{3\pi}{2}+\phi_m\right)\\
    \hat{U}_{z\text{CD}}(\alpha) &= \hat{\mathcal{D}}(\hat\sigma_z\alpha) = \hat{S}_y^\dagger\hat{U}_{x\text{CD}}(\alpha) \hat{S}_y,
\end{align}
    where $\hat{S}_y=R(\pi/2,\pi/2)$.

\subsection{Measure of the Characteristic Function of the Motional Wavepacket}\label{sec:motional-tomography}

\begin{figure}[b]
    \centering
    \begin{tikzpicture}
        \draw (0,0) node[xshift=0,yshift=-2,left] {$\ket{g}$} -- (5,0);
        \draw[decorate, decoration={snake, amplitude=0.2mm}] (0,-1) node[xshift=0,yshift=-2,left] {$\ket{\psi_m}$} -- (5,-1);
        \node[draw=black, fill=white] at (0.9,0) {$R_x(\theta)$};
        \def\x{1};
        \draw (5,\x/2) --++ (\x,0) --++ (0,-\x) --++ (-\x,0) -- cycle;
        \node at (5+\x/2,-\x/5) {$Z$};
        \draw (5.88,0.12) arc (45:135:0.5) (5.6,0.12) --++ (0.2,0.2);
        \draw (3, -1) --++ (0, 1) node {$\blacklozenge$};
        \node[draw=black, fill=white] at (3,-1) {$\hat{\mathcal{D}}(\sigma_x\beta/2)$};
    \end{tikzpicture}
    \caption{The circuit used to probe the characteristic function $\chi(\beta)$ of the motional state $\ket{\psi_m}$ by way of the conditional displacement gate $\hat{\mathcal{D}}(\sigma_x\beta/2)$ on a probe qubit.}
    \label{fig:sdf-readout}
\end{figure}

    Motional tomography can be performed by reading out the characterstic function in motional phase space using the conditional displacement gate on a probe qubit prepared in the $\ket{g}$ state~\cite{fluhmann-readout}. The probe qubit has the following operation applied
\begin{equation}
    \hat{\mathcal{D}}(\pm\sigma_x\beta/2)R_x(\theta)
\end{equation}
    before being measured in the $Z$-basis, as seen in \Cref{fig:sdf-readout}. The expectation value for the concluding $Z$ measurement on the probe qubit is related to the characteristic function in the following way
\begin{equation}
    -\braket{\sigma_z} = 1-2P = \cos\theta\; \mathfrak{Re}[\chi(\beta)] + \sin\theta\; \mathfrak{Im}[\chi(\beta)],
\end{equation}
    where $\theta$ is the angle applied to the $R_x(\theta)$ gate used on the qubit, $\beta$ is a complex-valued point in the motional phase space, and $\chi$ is the characteristic function:
\begin{equation}
    \chi(\beta) = \braket{\mathcal{D}(\beta)}.
\end{equation}
    Therefore, when $\theta=\pi/2$ is used, the expectation value of the probe qubit, measured in the $z$-basis, corresponds to $\mathfrak{Im}[\chi]$ and when no rotation is applied, the expectation value corresponds to $\mathfrak{Re}[\chi]$. We provide a detailed derivation in Appendix\ (\ref{app: chi_derivation}).

\begin{figure}[htb]
    \centering
    \includegraphics{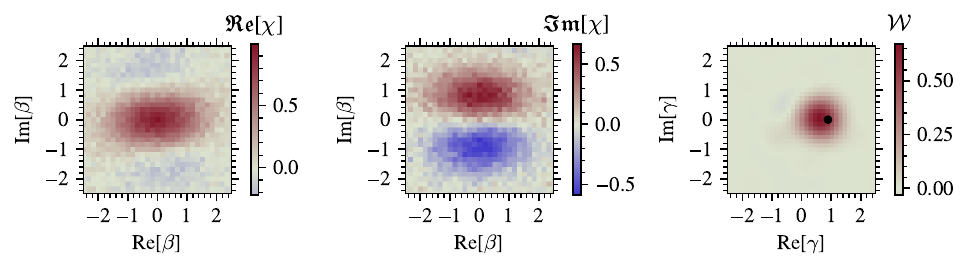}
    \caption{A coherent state prepared with a target displacement of $\alpha=0.89$ using $\hat{U}_\text{zCD}(-\alpha)$ on $\ket{g,0}$. The real and imaginary parts of $\chi(\beta)$ are read out using the circuit described in \Cref{fig:coherent-readout}, and $\mathcal{W}$ is calculated from the data using \Cref{eq:wigner}. The black dot on the Wigner representation is placed at $\gamma=0.89$. Conditional displacements are not an error free process, and both the CD used for preparing the coherent state and the CD used for reading out the state contribute to the fact that the probability density in the Wigner plot is not centered around $\gamma=0.89$.}
    \label{fig:coherent-readout}
\end{figure}

    To readout $\chi(\beta)$ on a circuit, values of complex $\beta$ are scanned using $\theta=0$ as well as $\theta=\pi/2$. This data can then be plotted as a colormap with the $x$-axis representing $\mathfrak{Re}[\beta]$ and the $y$-axis representing $\mathfrak{Im}[\beta]$ (see \Cref{fig:coherent-readout} for an example).

    The characteristic function is closely related to the Wigner representation, which is a psuedo-probability density of the motional wavepacket:
\begin{equation}
    \label{eq:wigner}
    \mathcal{W}(\gamma) = \frac{1}{\pi^2}\int\chi(\beta)e^{\gamma\beta^*-\gamma^*\beta}\;d^2\beta,
\end{equation}
    where $\gamma$, like $\beta$ is a complex-valued position in phase space. 
    This is a 2D Fourier transform that does an inverse Fourier transform from $\mathfrak{Re}[\beta]$ to $\mathfrak{Im}[\gamma]$ and a forward Fourier transform from $\mathfrak{Im}[\beta]$ to $\mathfrak{Re}[\gamma]$.

    The characteristic function provides a direct way to estimate the fidelity of a motional state. Through an expansion of the motional density matrix in terms of displacement operators, we can show that the fidelity of the motional state is computed by an overlap integral with a target motional state's characteristic function $\tilde{\chi}(\beta)$. Converting an integral over continuous $\beta$ to a sum over discretized $\beta$, we estimate the fidelity via 
    \begin{equation}
        \mathcal{F} = \frac{1}{\pi} \sum_{\beta} (\Delta \beta)^2 \tilde{\chi}(-\beta) \chi(\beta),
    \end{equation}
    \noindent{where} the sum is performed over a square discretized grid of $\beta$ values with discretization $\Delta \beta$.

\section{Conditional Rotation Gate}\label{sec:cr}
    The conditional rotation gate accumulates phase on the qubit conditioned on the motional state. The definition of the gate is
\begin{equation}
    \hat{U}_\text{CR}(\theta) = \exp\left[
        -i\frac{\theta}{2}\hat{\sigma}_z \hat{a}^\dagger \hat{a}
    \right] = \cos\left(\frac{\hat{N}\theta}{2}\right)\identity -i \sin\left(\frac{\hat{N}\theta}{2}\right)\hat{\sigma}_z,
\end{equation}
    \noindent{where} $\hat{N} = \hat{a}^{\dagger}\hat{a}$ is the number operator for the target qumode of the conditional rotation. Therefore, it is clear that this gate accumulates phase on the qubit at a rate dependent on the populations in the motional state.

    To implement this gate, we leverage light shifts accumulated from sequential detuned red and blue sideband pulses. The resulting dynamics lead to a phonon number $n$ dependence in the accumulated light shift. We implement the following protocol: (1) Apply a red sideband pulse (\Cref{eq:rsb-h}) of duration $\tau$ with blue detuning $\delta$, followed by a (2) blue sideband pulse (\Cref{eq:bsb-h}) of the same duration $\tau$ and blue detuning $\delta$. To illustrate how this leads to a conditional rotation, we consider each sideband Hamiltonian and include the carrier contribution: $\hat{H} = \hat{H}_{K} + \hat{H}_{C}$. Here, we denote the detuning of the carrier pulse as $\Delta_{K} = \omega_{\mu,m} + \delta$. Each Hamiltonian has the form $\hat{H} = \hat{h}e^{-i \omega t} + \text{h.c.}$, and the real-time propagator can be constructed via a Magnus approach~\cite{sutherland2021, bazavan_squeezing_2026}: 
\begin{equation}
    \hat{U} = \exp \left(\sum_{p=1}^{\infty} \hat{M}_{p}(t) \right),
    \label{eq: Magnus}
\end{equation}
    where $M_{p}(t)$ denotes the $p^{\rm th}$ order Magnus expansion contribution. The first-order term represents a displacement process $\hat{M}_{1}(t) = (-i/\hbar) \int_{0}^{t} d\tau\ \hat{H}(\tau)$, which will contribute negligibly for laser tones that are sufficiently detuned from sideband or carrier resonance. For second-order processes, the commutator is $[\hat{H}(t_{1}), \hat{H}(t_{2})] = -2i \sin(\omega(t_{1} - t_{2}))[\hat{h}, \hat{h}^{\dagger}]$. Therefore, the second-order Magnus approach can be time-integrated exactly: 
\begin{equation}
    \hat{M}_{2}(t) = -\frac{1}{2 \hbar^2} \int_{0}^{t} d\tau \int_{0}^{\tau} d\tau' [\hat{H}(\tau), \hat{H}(\tau')] = i[\hat{h}, \hat{h}^{\dagger}]\left (\frac{t}{\omega} - \frac{\sin(\omega \tau)}{\omega^2} \right ),
\end{equation}
    where the secular portion is proportional to time $t$ and the fast oscillatory term can be ignored for sufficient detunings $\omega \propto \Delta_{K}$ or $\delta$. A time-independent, effective Hamiltonian in the interaction picture is thus $\hat{H}_{\rm eff}(t) = [\hat{h}, \hat{h}^{\dagger}]/\omega$. The first-order Magnus term is $M_{1} \sim ||h||/\omega$, which decays algebraically with the frequency and can be neglected for detuned processes. 

    For each sideband pulse, the Hamiltonian can be written as $\hat{H}_{K} = \sum_{\ell} \hat{h}_{\ell} e^{-i\omega_{\ell}t} + \text{h.c.}$.  Therefore, the effective Hamiltonian for each sideband pulse is a sum over contributions by linearity of the Magnus expansion and integration operations~\cite{james_effective_2007}:
\begin{equation}
    \hat{H}_{\rm eff} = \sum_{\ell} \frac{[\hat{h}^{\dagger}_{\ell}, \hat{h}_{\ell}]}{\omega_{\ell}}.
\end{equation}
    For the sideband Hamiltonians for a target mode $m$ from manifold $\mu$, the static operators are $\hat{h}_{R} = (\Omega\eta /2) \sigma_{+} \hat{a}$ and $\hat{h}_{B} = (\Omega\eta /2) \sigma_{+} \hat{a}^{\dagger}$. Therefore, the relevant commutator is $[\hat{h}^{\dagger}, \hat{h}] = -(\Omega\eta  /2)^2 (\sigma_{z} (\hat{N} + 1/2) \pm 1/2)$ with frequency $\delta$, where the $+$ ($-$) sign is taken for the red (blue) sideband process. For the carrier process, the relevant commutator is $[\hat{h}^{\dagger}, \hat{h}] = -(\Omega^2 /4) \sigma_{z}$ at frequency $\Delta_{R}$ ($\Delta_{B}$) for the red (blue) sideband step. This leads to the second-order derived effective Hamiltonians for the red and blue sideband steps:
\begin{equation}
    \frac{\hat{H}_{\rm eff, K}}{\hbar} = -\frac{(\Omega\eta  )^2}{4\delta} \left (\sigma_{z} \left(\hat{N} + \frac{1}{2}\right) \pm \frac{1}{2} \right) - \frac{\Omega^2}{4 \Delta_{K}} \sigma_{z},
\end{equation}
    where $K=R,B$ to denote the red or blue detuned sideband step. Then, the red-blue sequence can be constructed via combining the real-time propagators $\hat{U} 
    = \exp (-i \tau (\hat{H}_{\rm eff, B} + \hat{H}_{\rm eff, R}) )$. The carrier detuning contributions from the red and blue sideband operations nearly cancel: $\Delta^{-1}_{R} + \Delta^{-1}_{B} = -2\delta/(\omega^2_{\mu, m} - \delta^2)$, which is strongly suppressed by the MHz motional frequency $\omega_{\mu, m} \gg \delta$. The sum of the Hamiltonians is 
\begin{equation}
\begin{split}
    \frac{\hat{H}_{\rm eff, B} + \hat{H}_{\rm eff, R}}{\hbar} =& -\frac{(\Omega\eta  )^2}{2\delta} \sigma_{z} \left(\hat{N} + \frac{1}{2}\right) + \frac{\Omega^2 \delta}{2(\omega^2_{\mu, m} - \delta^2 )}\sigma_{z} \\
    =& -\frac{(\Omega\eta  )^2}{2\delta} \sigma_{z} \hat{N} + \epsilon \sigma_{z},
\end{split}
\end{equation}
    where we have lumped together the qubit-only light shift rate into $\epsilon \equiv \epsilon_{C} - (\Omega\eta )^2/(4\delta)$ with the carrier light shift rate $\epsilon_{C} \equiv \Omega^2 \delta / (2 (\omega^2_{\mu, m} - \delta^2))$. The existence of this qubit-only light shift counterbalances the gate's target operation: A higher detuning leads to better-posed second-order dynamics while increasing the magnitude of $\epsilon_{C}$. Fortunately, the detuning can be chosen to \emph{cancel} these qubit-only light-shift contributions: $\delta^* = \eta  \omega_{\mu, m} / \sqrt{2 - \eta^2_{\mu, m}} \approx \eta \omega_{\mu, m} / \sqrt{2}$. As a result, the real-time propagator of the detuned red-blue sideband sequence includes both the target controlled rotation term and a residual qubit-only term: 
    \begin{equation}
        \hat{U}(t) = \exp \left( i \frac{(\Omega\eta  )^2 t}{2\delta} \sigma_{z} \hat{N} - i \epsilon t \sigma_{z}  \right ).
    \end{equation}
    
    The conditional rotation angle achieved is thus $\theta = (\Omega\eta )^2 / (2\delta)$. Since the lightshift causes the qubit to accumulate phase, it effectively acts as a $\hat{R}_z$ gate with an explicit dependence on the motional state: 
\begin{equation}
    \label{eq:cr}
    \hat{U}_\text{CR}(\theta) = \hat{R}_z(\theta=\hat{N}\Omega^2\eta^2  t/\delta) = \exp\left[
        i\frac{\eta^2\Omega^2 t}{2\delta}\hat\sigma_z\hat{N}
    \right].
\end{equation}

\begin{figure}[htb] 
\centering
\includegraphics{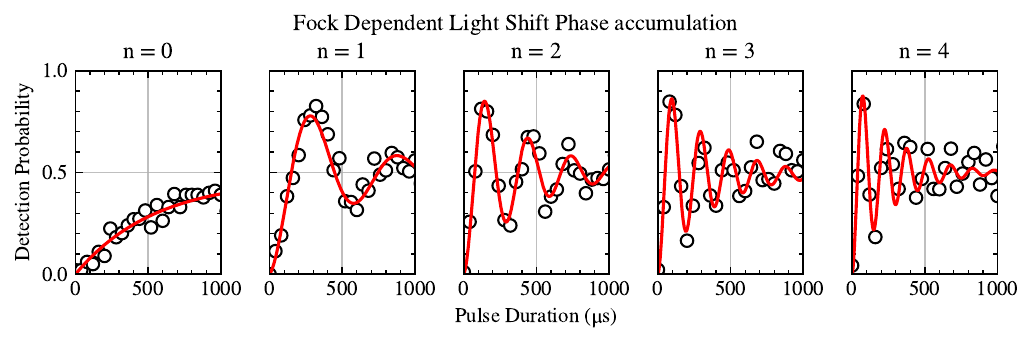}
    \caption{Results of a $\hat{S}_y^\dagger\hat{U}_\text{CR}(\theta=\hat{N}\Omega^2\eta^2 t/\delta)\hat{S}_y$ circuit with varying time on the CR gate on different Fock state initial states. It can be seen that for higher initial Fock states, the phase accumulation rate increases. However, the motional-phase decoherence is very dominant in this process.}
    \label{fig:lightshift-rabi}
\end{figure}

\begin{figure}[htb] 
\centering
\includegraphics{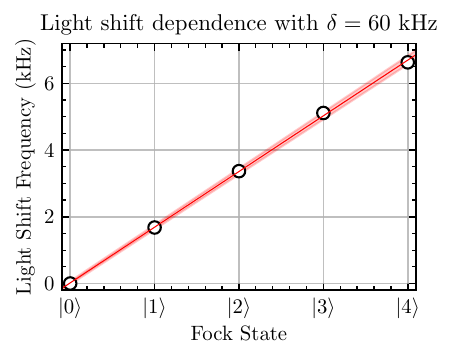}
    \caption{The light shift frequency (the frequency at which the qubit will accumulate phase) is linear with $n$.}
    \label{fig:lightshift}
\end{figure}


\section{Four-Tone Gates}\label{sec:gates-prog}
Our trapped-ion implementation provides access to a host of Gaussian operations via interference of first-order sideband processes. Following the approach of Sutherland and Srinivas~\cite{sutherland2021}, we implement a qubit-state dependent, or  conditional, beamsplitter and squeezing operation using two simultaneous spin-dependent forces along orthogonal spin axes. The non-commuting nature of the combined orthogonal spin-dependent forces lead to an effective higher-order motional operation conditioned on the mutually orthogonal spin axis. 

For both of the following gate protocols, the spin phase of one of the red-blue sideband pairs is set to generate a $\sigma_{x}$ interaction, while the the spin phase of the other red-blue sideband pair generates a $\sigma_{y}$ interaction. The detuned four tones $\omega_{L,k}$ (where $k$ indexes the four tones) result in a dominant two-photon $\sigma_{z}$ process that gives either a spin-dependent beamsplitter or spin-dependent squeezing operation, depending on the tones' detuning scheme.
    
Although multiple ions are present, we consider a model for the two pairs of first-order red and blue sideband tones on a target ion. One pair of red and blue sideband tones targets a $\sigma_{x}$ interaction on mode $m$ from manifold $\mu$, while another pair of red and blue sideband tones targets a $\sigma_{y}$ interaction on mode $m'$ from manifold $\mu'$. This leads to the interaction Hamiltonian from two SDFs $\hat{H}_{\rm SDF}(\phi_{q} = 0, \phi_{m} = 0) + \hat{H}_{\rm SDF}(\phi_{q} = \pi/2, \phi_{m} \equiv \phi)$ from \Cref{eq: SDF}: 
\begin{equation}
    \frac{\hat{H}(t)}{\hbar} = \frac{\Omega(t)}{2i} \left [ \eta_{\mu, m} \sigma_{x} (\hat{a}_{\mu, m} e^{-i \delta t} - \hat{a}^{\dagger}_{\mu, m} e^{i \delta t} ) + \eta_{\mu', m'}\sigma_{y}(\hat{a}_{\mu', m'} e^{-i n\delta t}e^{-i\phi} - \hat{a}^{\dagger}_{\mu', m'} e^{in\delta t}e^{i\phi} )  \right ],   
    \label{eq: 4tone_hamiltonian}
\end{equation}
\noindent{where} the two Lamb-Dicke factors are related to the same ion interacting with two different qumodes, the bosonic operators obey the standard Bose commutation relations for different motional modes, and we keep the assumption that both sideband pairs share the same time-dependent, real-valued Rabi rate $\Omega(t) \in \mathbb{R}$, whose time-dependence may be inherited from laser amplitude pulse-shaping. $\eta_{\mu, m}$ is the Lamb-Dicke parameter for the target ion with respect to motional mode $m$ from manifold $\mu$. The parameter $n=-1$ ($n=1$) sets the detuning relationships and produces a spin-dependent squeezing (beamsplitter) operation. The motional phase on the $\sigma_{y}$ tones sets the value of $\phi$. 

We invoke the Magnus expansion from \Cref{eq: Magnus} to understand the dominant processes at second order. As in the condition rotation gate, the dominant contribution for $n=\pm1$ occurs at second-order in $\hat{M}_{2}$ and corresponds to two-photon interference, where the commutators of the two spin dependent forces lead to the desired conditional operations. For the conditional beamsplitter and squeezing operations, it is precisely $\hat{M}_{2}$ which dominates the propagator in the limit $( \Omega\eta/\delta) \ll 1$. 

\subsection{Conditional Beamsplitter}\label{sec:cb}
    To generate the conditional beamsplitter (CBS) interaction~\cite{gan_hybrid_2020}, the four tones are set up with symmetric detuning, $\delta$, off of the two modes of interest (denoted with $\omega_{\mu,m}$ and $\omega_{\mu',m'}$), as follows:
\begin{align}
    \omega_{L,r} &= \omega_c - \omega_{\mu,m}+\delta\\
    \omega_{L,b} &= \omega_c + \omega_{\mu,m}-\delta\\
    \omega_{L,r'} &= \omega_c - \omega_{\mu',m'}+\delta\\
    \omega_{L,b'} &= \omega_c + \omega_{\mu',m'}-\delta,
\end{align}
    \noindent{where} $\omega_{L, k}$ refers to the laser tone with $k = r,b$ referring to targeting a red ($r$) or blue ($b$) sideband process. This detuning scheme corresponds to $n=+1$ in the parametrized 4-tone Hamiltonian (\Cref{eq: 4tone_hamiltonian}), which leads to a spin-dependent beamsplitter operation as the dominant Magnus expansion contribution. Analysis of the dominant two-photon processes in the Magnus expansion produce an estimate for the accrued beamsplitter angle:
\begin{equation}
    \theta(t) = \eta_{\mu, m} \eta_{\mu', m'} \int_{0}^{t} d\tau_{1} \int_{0}^{\tau_{1}} d\tau_{2}\ \Omega(\tau_{1}) \Omega(\tau_{2}) \sin (\delta (\tau_{2} - \tau_{1})),
    \label{eq: CBS_angle}
\end{equation}
    \noindent{where}, for example, $\theta = \pi$ swaps the Fock occupancy of one mode with the occupancy of another mode. After redefining the $\sigma_{y}$ motional phase from \Cref{eq: 4tone_hamiltonian} as $\phi \to \phi + \pi/2$, the $\hat{M}_{2}$ Magnus contribution will correspond to a spin-dependent beamsplitter unitary of the form 
\begin{equation}
    \hat{U}_\text{CBS}(\theta,\phi) = \exp\left[
        \frac{i\theta(t)}{2}\sigma_z(a^\dagger b \;e^{-i\phi} + ab^\dagger e^{i\phi})
    \right], \label{eq: controlled_beamsplitter}
\end{equation}
    where $a$ ($b$) and $a^\dagger$ ($b^{\dagger}$) are the annihilation and creation operators for the first (second) mode described by frequency $\omega_{\mu,m}$ ($\omega_{\mu',m'}$). The unitary described by \Cref{eq: controlled_beamsplitter} omits a global phase proportional to $\exp(i \theta(t) I_{d})$, where $I_{d}$ is the identity operator on the full qubit-motional Hilbert space. A particular value of $\theta$ is achieved by designing the protocol described by an amplitude shape $\Omega(t)$ and detuning $\delta$ to achieve a $\theta$ by the end of the protocol via \Cref{eq: CBS_angle}. The phase of the gate $\phi$ is controlled by setting the motional phase on one of the sideband pairs to $\phi$ while setting the other sideband pairs' motional phase to zero. 

    To demonstrate the conditional beamsplitter, we prepare a Fock state of $\ket{1}$ in qumode $\mu, m$ and $\ket{0}$ in qumode $\mu',m'$, and then the beamsplitter is applied for variable time (see \Cref{fig:beamsplitter-circ}). After the beamsplitter gate is applied, the state becomes
\begin{equation}
    \hat{U}_\text{CBS}(t)\ket{g,1,0}=\cos(\Omega_\text{eff}t)\ket{g,1,0}+\sin(\Omega_\text{eff}t)\ket{g,0,1},
\end{equation}
    where $\Omega_\text{eff}$ is the effective Rabi rate of the operation which is determined within \Cref{eq: CBS_angle}. To read out the motional state of each mode, a JC gate is applied with $\theta=\pi$ to effectively project the motional state onto the qubit state such that $\hat{U}_\text{JC}(\pi)\ket{g,0} = \ket{g,0}$ and  
    $\hat{U}_\text{JC}(\pi)\ket{g,1} = \ket{e,0}$.
    The circuit is repeated both with applying the JC either to qumode $\mu,m$ and to $\mu',m'$ to read out the probability of $\ket{1}$ for both qumodes (see \Cref{fig:beamsplitter}).

\begin{figure}[htb] 
\centering
\begin{tikzpicture}
        \draw (0,0) node[xshift=0,yshift=-2,left] {$\ket{g}$} -- (5,0);
        \draw (0,-1) node[xshift=0,yshift=-2,left] {$\ket{g}$} -- (5,-1);
        \draw[decorate, decoration={snake, amplitude=0.2mm}] (0,-2) node[xshift=0,yshift=-2,left] {$\ket{0}$} -- (5,-2);
        \draw[decorate, decoration={snake, amplitude=0.2mm}] (0,-3) node[xshift=0,yshift=-2,left] {$\ket{0}$} -- (5,-3);
        
        \draw (0.9,-2) node [draw=black, fill=white] {$\text{AJC}(\pi)$} --++ (0, 1) node {$\blacklozenge$};
        
        \def\x{1};
        \draw (5,\x/2) --++ (\x,0) --++ (0,-\x) --++ (-\x,0) -- cycle;
        \node at (5+\x/2,-\x/5) {$Z$};
        \draw (5.88,0.12) arc (45:135:0.5) (5.6,0.12) --++ (0.2,0.2);

        \draw (5,\x/2-1) --++ (\x,0) --++ (0,-\x) --++ (-\x,0) -- cycle;
        \node at (5+\x/2,-\x/5-1) {$Z$};
        \draw (5.88,0.12-1) arc (45:135:0.5) (5.6,-1+.12) --++ (0.2,0.2);
        
        \draw (2.65,0) node {$\blacklozenge$} --++ (0,-2.5);
        \draw[fill=white] (2,.3-2) --++ (1.2,0) --++ (0,-1-.3*2) --++ (-1.2,0) -- cycle;
        \draw (2.6,-2.5) node {$\text{CBS}(t)$};

        \draw[blue] (4.5,-3) node [draw=blue, fill=white] {$\text{JC}(\pi)$} --++ (0, 3) node {$\blacklozenge$};
        
        \draw (4.2,-2) node [draw=black, fill=white] {$\text{JC}(\pi)$} --++ (0, 2) node {$\blacklozenge$};
        \node at (4.2, -2.5) {or};
    \end{tikzpicture}
    \caption{The circuit used to demonstrate the beamsplitter gate. The AJC prepares a Fock state of $\ket{1}$ on the first qumode $\mu,m$, and then the beamsplitter entangles the first qumode with the second qumode $\mu',m'$. A final JC gate is applied to either the first or second qumode to read out the probability of being in $\ket{0}$ or $\ket{1}$. Note: only one of the JC gates is applied each time the circuit is run. The data colors in \cref{fig:beamsplitter} correspond to the colors of the JC gates here.}
    \label{fig:beamsplitter-circ}
\end{figure}

\begin{figure}[htb] 
\centering
\includegraphics{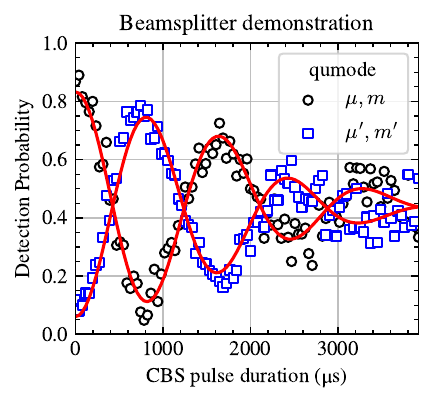}
    \caption{Results of reading out motional state probability as function of applying variable time on the beamsplitter operation. The circuit used for demonstration is shown in \Cref{fig:beamsplitter-circ}. The two possible motional states are $\ket{0}$ and $\ket{1}$, and these are projected onto the qubit state using the JC gate. Because the JC gate will not drive the $\ket{g,0}$ state, the qubit will only rotate to the bright state $\ket{e}$ when the motional state is $\ket{1}$, so in this case $P$ corresponds to the probability of being in the $\ket{1}$ state. The circuit is ran both with projecting qumode $\mu,m$ onto the qubit, and then again with projecting qumode $\mu',m'$ onto the qubit to get both curves.}
    \label{fig:beamsplitter}
\end{figure}

\subsection{Conditional Squeeze}\label{sec:cs}
    Qubit-dependent (conditional) squeezing (CSq) generates single-mode squeezing conditioned on the ion's internal state. Conditional squeezing operations have gained substantial interest in the community for generating multi-qubit entangling operations~\cite{katz_n-body_2022,katz_demonstration_2023}. However, generating conditional squeezing via second motional sidebands suffers from speed limitations with an effective rate $\sim \Omega\eta^2 $ and additionally requires pulse control requirements. Instead, the four-tone protocol described by \Cref{eq: 4tone_hamiltonian} provides a faster alternative to generating conditional squeezing and does not require sophisticated pulse control. To generate the conditional single-mode squeeze operation, the four tones are detuned by $\delta$ above and below the mode of interest with motional sideband frequency $\omega_{\mu,m}$, corresponding to the choice of $n=-1$ in \Cref{eq: 4tone_hamiltonian}:
\begin{align}
    \omega_{L,r-} &= \omega_c - \omega_{\mu,m}+\delta\\
    \omega_{L,b+} &= \omega_c - \omega_{\mu,m}-\delta\\
    \omega_{L,b-} &= \omega_c + \omega_{\mu,m}-\delta\\
    \omega_{L,b+} &= \omega_c + \omega_{\mu,m}+\delta.
\end{align} 
    The achieved complex-valued squeezing magnitude $\zeta$ is obtained by analysis of the second-order Magnus contribution: 
\begin{equation}
    \zeta(t) = \eta^2 e^{i \phi} \int_{0}^{t} d\tau_{1} \int_{0}^{\tau_{1}} d\tau_{2}\ \Omega(\tau_{1}) \Omega(\tau_{2}) \sin (\delta (\tau_{1} - \tau_{2}))  \equiv r(t)e^{i \phi} , \label{eq: squeezing_condition}
\end{equation}
    \noindent{where} $r(t)$ is the magnitude of the conditional squeezing accrued over the duration $t$ for the target mode $m$ of manifold $\mu$.
    It is evident from \Cref{eq: squeezing_condition} that the amplitude pulse-shape, pulse timing, Lamb-Dicke parameter, and detuning influence the achieved squeezing. A target squeezing phase $\phi$ is achieved by setting the motional phase on one of the sideband pairs to $\phi$ while setting the other sideband pair's motional phase to zero. Therefore, this scheme recovers the conditional squeezing operation:
\begin{equation}
    \label{eq:csq}
    \hat{U}_\text{CSq}(\zeta) = \exp\left[\frac{\sigma_{z}}{2}\left(\zeta^* a^2 - \zeta a^{\dagger 2}\right)\right].
\end{equation}

\begin{figure}[htb] 
\centering
\includegraphics{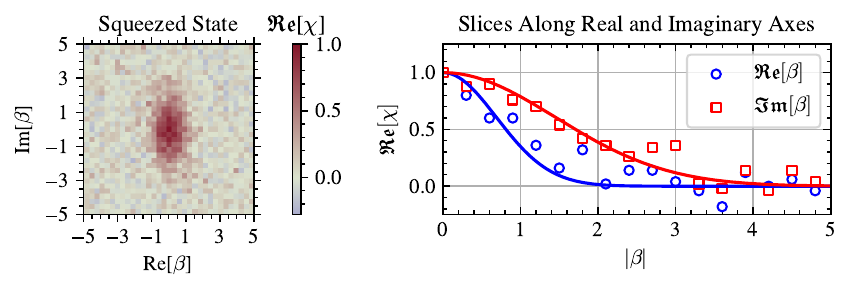}
    \caption{Example of the CSq being used to prepare a squeezed state, with the characteristic function read out using the circuit described in \Cref{fig:sdf-readout}. The slices along the real and imaginary axes show the uncertainty in $\mathfrak{Re}[\beta]$ decreasing while the uncertainty in $\mathfrak{Im}[\beta]$ increasing. A fit of the data yields a squeezing parameter $r=0.37$.}
    \label{fig:squeeze}
\end{figure}

    An example of applying the CSq gate to prepare a squeezed vacuum state is shown in \Cref{fig:squeeze}.

\section{Conclusions}\label{conclusion}
    We have demonstrated an optical implementation of a universal gate set for hybrid qubit-boson quantum computation on a trapped-ion quantum processor. Leveraging primarily first-motional-sideband interactions and qubit-carrier transitions, we describe how our implementation achieves several unitaries of interest for quantum simulation and computation.   
    When combined with our single-qubit rotations and M{\o}lmer-S{\o}rensen gates \cite{qscout}, our implementation of the (anti-)Jaynes-Cummings, conditional displacement, conditional rotation, conditional beamsplitter, and conditional squeeze operations constitute a universal gate set for hybrid CV-DV computing. Our demonstrations show achievement of these unitaries, including discussion of error from ideality. This work has grown out of recent interest to utilize QSCOUT for CV-DV related research~\cite{triggate,yukawa,lim2026,snu,hybridlane,hyqbench}.

    Realizing a universal gate set for CV-DV computing provides a path forward towards more complex demonstrations of hybrid qubit-boson computing. In principle, our gate set provides access to fault-tolerant encodings of trapped-ion qubit information via motional states such as Gottesman–Kitaev–Preskill grid states \cite{gottesman_encoding_2001} or squeezed cat states \cite{kiryu_linear_2025}. The coherent operations demonstrated here serve as primitives for more sophisticated protocols built from compositions of each gate. For example, non-Gaussian motional operations \cite{triggate} or many-qubit operations \cite{katz_n-body_2022} may be synthesized from combinations of primitives in our gate set, and, as such, may directly benefit the fields of boson-only and qubit-only quantum computing, respectively. 

\section{Acknowledgments}

    We acknowledge the various contributions of our user teams in the joint development of this gateset for QSCOUT including: 

    Mahn-Soo Choi, Wonhyeong Choi, Anurag Dwivedi, Jim Furches, Matt Grau, James Graves, Chanyang Im, Srinivasan Iyengar, Jiyong Kang, Junki Kim, Taehyun Kim, Sungjoo Lim, Yuan Liu, Noah Lordi, Peter Love, Shubdeep Mohapatra, Jake Montgomery, Tommaso Rainaldi, Phil Richerme, Felix Ringer, George Siopsis, Jieqiu Shao, Qingfeng Wang, and Huiyang Zhou.
    
    Sandia National Laboratories is a multi-mission laboratory managed and operated by National Technology \& Engineering Solutions of Sandia, LLC (NTESS), a wholly owned subsidiary of Honeywell International Inc., for the U.S. Department of Energy’s National Nuclear Security Administration (DOE/NNSA) under contract DE-NA0003525. This written work is authored by an employee of NTESS. The employee, not NTESS, owns the right, title and interest in and to the written work and is responsible for its contents. Any subjective views or opinions that might be expressed in the written work do not necessarily represent the views of the U.S. Government. The publisher acknowledges that the U.S. Government retains a non-exclusive, paid-up, irrevocable, world-wide license to publish or reproduce the published form of this written work or allow others to do so, for U.S. Government purposes. The DOE will provide public access to results of federally sponsored research in accordance with the DOE Public Access Plan. SAND2026-29513O.

\begin{appendices}
\section{Conventions}\label{app: conventions}

    In quantum optics literature, $\sigma_z$ is defined as in \Cref{eq:z-convention}, such that the higher (lower) energy state $\ket{e}$ ($\ket{g}$) is the $+1$ ($-1$) eigenstate of $\sigma_z$, and in quantum computing literature, $\sigma_z=\proj{0}-\proj{1}$, such that $\ket{0}$ ($\ket{1}$) is chosen to be the $+1$ ($-1$) eigenstate of $\sigma_z$. In trapped-ion literature, depending on the ion and the qubit encoding, sometimes the lower energy state $\ket{g}$ will be labeled either $\ket{0}$ or $\ket{1}$. Conventionally, $^{171}$Yb ions label the lower energy state $\ket{S_{1/2},F=0,m=0}\equiv\ket{g}:=\ket{0}$ and the higher energy state $\ket{S_{1/2},F=1,m=0}\equiv\ket{e}:=\ket{1}$, such that $\ket{g}$ is the +1 eigenstate under quantum computing convention and the -1 eigenstate under quantum optics convention. Here we discuss the consequences and remedies for using ions where the lower energy state is labeled $\ket{0}$ for hybrid CV-DV quantum computing.

    When the lower energy state $\ket{g}$ is labeled $\ket{0}$, then the definition of the single-qubit $z$ Pauli Matrix is $\sigma_z=\proj{0}-\proj{1}=\proj{g}-\proj{e}$, which is in direct contradiction to the definition in \Cref{eq:z-convention}. To distinguish this definition of $\sigma_z$ and the one used in this paper, we define
\begin{align}
    \sigma_z^{(q)} &= \proj{g}-\proj{e}\\
    \sigma_z^{(m)} &= \proj{e}-\proj{g},
\end{align}
    where $\sigma_z^{(q)}$ describes the convention used for qubit-only quantum computing when the lower energy state is labeled $\ket{0}$ and $\sigma_z^{(m)}$ describes the convention used for hybrid qubit-motion quantum computing and in this paper (see equation\ (\ref{eq:z-convention} from the main text).
    
    The two conventions can be related by a global phase and correspond to a unitary transformation of each single-qubit Pauli operator for $N$ qubits: $U = \sigma_x^{\otimes N} \equiv \bigotimes_{i=1}^{N} \sigma_x^{(i)} = \sigma_x^{(1)} \otimes \sigma_x^{(2)} \otimes \dots \otimes \sigma_x^{(N)}$ on each element $P \in \mathcal{P}_{N}$ of the $N$-qubit Pauli group $\mathcal{P}_{N} = \{\sigma_{x}, \sigma_{y}, \sigma_{z}\}^{\otimes N}$. Specifically, each element $P$ transforms as $P \to \sigma^{\otimes N}_{x} P \sigma^{\otimes N}_{x}$, which leads to a simple sign change in the definitions of the single-qubit Pauli operators:
\begin{equation}
    \sigma_z^{(q)}=-\sigma_z^{(m)}.
\end{equation}
\begin{equation}
    \sigma_x^{(q)}=\sigma_x^{(m)},
\end{equation}
\begin{equation}
    \sigma_y^{(q)}=-\sigma_y^{(m)}.
\end{equation}
    As such, both conventions leave $\sigma_{x}$ unchanged as $\sigma_x=\ketbra{e}{g}+\ketbra{g}{e}$.

\begin{figure}[t!]
    \centering
    \begin{tikzpicture}[]
    \coordinate (o) at (0, 0);
    \draw (o) ++ (-6,2.6) node[anchor=north west] {$\ket{e} \equiv \ket{S_{1/2},F=1,m=0}:$};
    \draw (o) ++ (-6,-2) node[anchor=north west] {$\ket{g} \equiv \ket{S_{1/2},F=0,m=0}:$};
    \draw [shift={(o)}, lightgray, fill, fill opacity=0.1] (o) --++ (56.7:0.4) arc (56.7:90.:0.4) -- cycle;
    \draw [shift={(o)}, lightgray, fill, fill opacity=0.1] (o) --++ (-135.7:0.4) arc (-135.7:-33.2:0.4) -- cycle;
    \draw(o) circle (2cm);
    \draw [rotate around={0.:(o)},dash pattern=on 3pt off 3pt] (o) ellipse (2cm and 0.9cm);
    \draw (o) --++ (0.70,1.07);
    \draw [-stealth] (o) --++ (0,2);
    \draw [-stealth] (o) --++ (-0.81,-0.79);
    \draw [-stealth] (o) --++ (2,0);
    \draw [dotted] (o) ++ (0.7,1) --++ (0,-1.46);
    \draw [dotted] (o) --++ (0.7,-0.46);
    \draw (o) ++ (-0.08,-0.3) node[anchor=north west] {$\phi$};
    \draw (o) ++ (0.01,0.9) node[anchor=north west] {$\theta$};
    \draw (o) ++ (-1.01,-0.72) node[anchor=north west] {$\mathbf {\hat{x}}$};
    \draw (o) ++ (2.07,0.3) node[anchor=north west] {$\mathbf {\hat{y}}$};
    \draw (o) ++ (-0.5,2.6) node[anchor=north west] {$\mathbf {\hat{z}}=\ket{0}=\ket{e}$};
    \draw (o) ++ (-0.4,-2) node[anchor=north west] {$-\mathbf {\hat{z}}=\ket{1}=\ket{g}$};
    \draw (o) ++ (0.4,1.65) node[anchor=north west] {$\ket{\psi}$};
    \draw [fill] (o) circle (1.5pt);
    \draw [fill] (o) ++ (0.7,1.1) circle (0.5pt);
    
    \coordinate (o) at (6, 0);
    \draw [shift={(o)}, lightgray, fill, fill opacity=0.1] (o) --++ (-90:0.4) arc (-90:-120.:0.4) -- cycle;
    \draw [shift={(o)}, lightgray, fill, fill opacity=0.1] (o) --++ ((-135.7:0.4) arc (-135.7:-162
    :0.4) -- cycle;
    \draw(o) circle (2cm);
    \draw [rotate around={0.:(o)},dash pattern=on 3pt off 3pt] (o) ellipse (2cm and 0.9cm);
    \draw [-stealth] (o) --++ (0,-2);
    \draw [-stealth] (o) --++ (-0.81,-0.79);
    \draw [-stealth] (o) --++ (-2,0);
    \draw [dotted] (o) --++ (-0.7,-0.21) --++ (0,-1.);
    \draw [fill] (o) --++ (-.7,-1.21) circle (0.5pt) node[anchor=north east, xshift=7] {$\ket{\psi}$};
    \draw (o) ++ (-0.75,-0.02) node[anchor=north west] {$\phi$};
    \draw (o) ++ (-.4,-0.45) node[anchor=north west] {$\theta$};
    \draw (o) ++ (-1.15,-.18) node[anchor=north west] {$\mathbf {\hat{x}}$};
    \draw (o) ++ (-2.07,0.3) node[anchor=north east] {$\mathbf {\hat{y}}$};
    \draw (o) ++ (-0.5,2.6) node[anchor=north west] {$-\mathbf {\hat{z}}=\ket{1}=\ket{e}$};
    \draw (o) ++ (-0.4,-2) node[anchor=north west] {$\mathbf {\hat{z}}=\ket{0}=\ket{g}$};
    \draw [fill] (o) circle (1.5pt);
    \end{tikzpicture}
    \caption{
        Bloch spheres relative to $^{171}$Yb states $\ket{e}$ and $\ket{g}$. Note that in each Bloch sphere, $\ket{0}$ ($\ket{1}$) labels the +1 (-1) eigenstate, as is conventional in quantum computing. (Left) Bloch sphere defined with $\sigma_z=\proj{e}-\proj{g}$, as in this manuscript. (Right) Bloch sphere defined with $\sigma_z=\proj{g}-\proj{e}$, which is used for qubit-only computing on our system. To translate between the two, it can be seen that the direction $\phi$ propagates from the $x$-axis must reverse.
    }
    \label{fig:blochs}
\end{figure}

    An important consequence of changing between these conventions is that the Pauli raising and lowering operators $\sigma_\pm$ have the opposite effect in each convention, as in
\begin{equation}
    \sigma_\pm^{(q)}=\frac{\sigma_x^{(q)}\pm\sigma_y^{(q)}}{2} = \frac{\sigma_x^{(m)}\mp\sigma_y^{(m)}}{2} = \sigma_\mp^{(m)}.
\end{equation}
    In other words, using $\sigma_z^{(q)}$ makes $\ket{g}$ the raised state, and $\ket{e}$ the lowered state. Therefore, if we kept our qubit-only convention, the Jaynes-Cummings and Anti-Jaynes-Cummings would be written with the ``wrong'' Pauli raising and lowering operators:
\begin{align}
    \hat{U}_\text{JC}^{(q)}(\theta, \phi) &= \exp\left[-i\frac{\theta}{2}( \hat{\sigma}_-^{(q)}\hat{a} e^{-i\phi}+\hat{\sigma}_+^{(q)}\hat{a}^\dagger e^{i\phi})\right] \\
    \hat{U}_\text{AJC}^{(q)}(\theta, \phi) &= \exp\left[-i\frac{\theta}{2}(\hat{\sigma}_-^{(q)}\hat{a}^\dagger e^{-i\phi}+ \hat{\sigma}_+^{(q)}\hat{a} e^{i\phi})\right].
\end{align}

    It can be seen in \Cref{fig:blochs} that when moving from one convention to the other, the direction of positive $\phi$ reverses. Therefore, to implement the $\sigma_z^{(m)}$ notation, that is used in this paper, we switch the direction that our $R_z$ gates are applied, as in
\begin{equation}
    R_z^{(m)}(\theta) = R_z^{(q)}(-\theta),
\end{equation}
    and we reverse the direction of the phase in $R$ gates, as in
\begin{equation}
    R^{(m)}(\theta,\phi) = R^{(q)}(\theta,-\phi).
\end{equation}


\section{Jaqal Gate Calls}

    Here we describe how to program the hybrid qubit-motion gates using Jaqal$^\text{TM}$. The qubit-only gates are described in \cite{jaqal}. All angles and phases are given in radians.

    \Cref{eq:JC}, and Anti-Jaynes-Cummings, \Cref{eq:AJC}, are called with
\begin{equation}
    \texttt{JC <qubit> <manifold> <mode> <phase> <angle>}
\end{equation}
    and
\begin{equation}
    \texttt{AJC <qubit> <manifold> <mode> <phase> <angle>}
\end{equation}
    respectively, where \texttt{<qubit>}, \texttt{<manifold>}, and \texttt{<mode>} are indices describing which qubit, manifold, and mode are to be addressed, \texttt{<phase>} is the phase of the gate $\phi$, and \texttt{<angle>} is the angle of the gate $\theta$. These gates can also be used to prepare a Fock state with
\begin{equation}
    \texttt{FockStatePrep <qubit> <manifold> <mode> <Fock\_number>}
\end{equation}
    where \texttt{<Fock\_number>} is the target Fock state.
    
    The conditional displacement can be called with control over qubit phase $\phi_q$, displacement phase $\phi_m$, and displacement amplitude $|\alpha|$ with
\begin{equation}
    \texttt{CD <qubit> <manifold> <mode> <phase\_q> <phase\_m> <alpha>}
\end{equation}
    Alternatively, a CD gate can be called specifically conditioned on the $x$, $y$, or $z$ basis with arguments for the real and imaginary part of $\alpha$
\begin{align}
    \texttt{xCD <qubit> <manifold> <mode> <re\_alpha> <im\_alpha>}\\
    \texttt{yCD <qubit> <manifold> <mode> <re\_alpha> <im\_alpha>}\\
    \texttt{zCD <qubit> <manifold> <mode> <re\_alpha> <im\_alpha>}
\end{align}
    The measurement circuit for reading out $\chi$, described by \Cref{fig:sdf-readout} can be called with
\begin{equation}
    \texttt{ChiMeas <qubit> <manifold> <mode> <re\_beta> <im\_beta> <re\_part>}
\end{equation}
    where \texttt{<re\_part>} is 1 for reading out $\mathfrak{Re}[\chi(\beta)]$ and 0 for reading out $\mathfrak{Im}[\chi(\beta)]$.

    For the following conditional gates (CR, CBS, CSq), the motional operation is conditioned on the qubit's $Z$-projection via $\sigma_{z}$. As a result, these gates do not require nor have a phase argument $\sigma_{q}$ via \texttt{<phase\_q>}, in contrast to the conditional displacement \texttt{CD} gate. Here, we detail the Jaqal calls for these motional gates conditioned on $\sigma_{z}$. 
    
    The conditional rotation, \Cref{eq:cr}, is called with 
\begin{equation}
    \texttt{CR <qubit> <manifold> <mode> <angle>}
\end{equation}
    where \texttt{<angle>} corresponds to $\theta$.

    The conditional beamsplitter, \Cref{eq: controlled_beamsplitter}, called with
\begin{equation}
    \texttt{CBS <qubit> <manifold\_1> <mode\_1> <manifold\_2> <mode\_2> <phase> <angle>}
\end{equation}
    where both modes involved in the beamsplitter must be defined, \texttt{<angle\_rad>} defines the rotation of the gate $\theta$ on the Bloch sphere (where an angle of $\pi$ corresponds to full swap of population between the modes), and \texttt{<phase>} is the beamsplitter phase $\phi$.

    The conditional squeeze, \Cref{eq:csq}, is called with
\begin{equation}
    \texttt{CSq <qubit> <manifold> <mode> <amp\_sq> <phase\_sq>}
\end{equation}
    where the amplitude \texttt{<amp\_sq>} and phase \texttt{<phase\_sq>} define the complex squeezing amplitude $\zeta$.


\section{Motional readout derivation} \label{app: chi_derivation}
    Here we show how a motional mode's state can be understood by measurement of a probe qubit. Following the procedure in \cite{fluhmann-readout}, the characteristic function $\chi(\beta)$ can be measured by performing conditional displacements on a probe qubit, followed by measurement of qubit's state. For this derivation, we use the $\sigma_{z} \equiv \sigma^{(m)}_{z}$ convention employed throughout this work, detailed in Appendix\ \ref{app: conventions} and in equation\ (\ref{eq:z-convention}).   
    The circuit in \Cref{fig:sdf-readout} applies the operation $\hat{\mathcal{D}}(\hat\sigma_x\beta/2)\hat{R}_x(\theta)$ to the state $\ket{g} \otimes \ket{\psi_m}$ to produce state $\ket{\Psi}$, such that $\langle \sigma_{z}\rangle \equiv \braket{\Psi|\hat\sigma_z|\Psi}$ determines the characteristic function. We write this expectation under the evolved state $\ket{\Psi} = \hat{\mathcal{D}}(\hat\sigma_x\beta/2)\hat{R}_x(\theta) \ket{g} \otimes \ket{\psi_m}$ and denote the initial combined qubit-motion state as $\ket{g, \psi}$:
\begin{equation}
\begin{split}
    \braket{\hat\sigma_z} &= \braket{g,\psi|\hat{R}^\dagger_x(\theta)\hat{\mathcal{D}}^\dagger(\hat\sigma_x\beta/2)\hat\sigma_z\hat{\mathcal{D}}(\hat\sigma_x\beta/2)\hat{R}_x(\theta)|g,\psi}\\
    &= \bra{g,\psi}\hat{R}^\dagger_x(\theta)
        \left[\proj{+}\hat{\mathcal{D}}(-\beta/2)+\proj{-}\hat{\mathcal{D}}(\beta/2)\right]
        \hat\sigma_z\left[\proj{+}\hat{\mathcal{D}}(\beta/2)+\proj{-}\hat{\mathcal{D}}(-\beta/2)\right]
    \hat{R}_x(\theta)\ket{g,\psi},
    \label{eq: sigma_z_char_function}
\end{split}
\end{equation}
\noindent{where} we have expanded the spin-dependent displacement operators in terms of the $\sigma_{x}$ spin eigenstates $\ket{\pm}$ from equation\ (\ref{eq: CD_spin_expansion}) with $\mathbf{\hat{n}} = \hat{x}$. We then evaluate the inner expectation values of the $\sigma_{z}$ operator with the $\ket{\pm}$ eigenstates using identities $\braket{\pm|\hat\sigma_z|\pm}=0$ and $\braket{\pm|\sigma_z|\mp}=1$ for matched and mismatched eigenstates, respectively. As a result, mismatched-eigenstate terms remain, simplifying \Cref{eq: sigma_z_char_function}: 
\begin{equation}
\begin{split}
    \braket{\hat\sigma_z} &= \bra{g,\psi}\hat{R}^\dagger_x(\theta)
        \left[\ketbra{+}{-}\hat{\mathcal{D}}(-\beta)+\ketbra{-}{+}\hat{\mathcal{D}}(\beta)\right]
    \hat{R}_x(\theta)\ket{g,\psi}.
\end{split}
\end{equation}
    We then invoke the definition of the single-qubit rotation operator from \Cref{eq:sqr} and $R^{\dagger}_{x} = \hat{R}(\theta = -\pi/2, \phi = 0)$ to simplify further via $\hat{R}^\dagger_x(\theta)\ket{\pm} = e^{\pm i\theta/2}\ket{\pm}$. By applying $\hat{R}^{\dagger}_{x}(\theta)$, we obtain 
\begin{equation}
\begin{split}
    \braket{\hat\sigma_z} &= \bra{g,\psi}
        \left[e^{i\theta}\ketbra{+}{-}\hat{\mathcal{D}}(-\beta)+e^{-i\theta}\ketbra{-}{+}\hat{\mathcal{D}}(\beta)\right]
    \ket{g,\psi}\\
    &= \bra{\psi}
        \left[e^{i\theta}\braket{g|+}\braket{-|g}\hat{\mathcal{D}}(-\beta)+e^{-i\theta}\braket{g|-}\braket{+|g}\hat{\mathcal{D}}(\beta)\right]
    \ket{\psi}.
    \end{split}
\end{equation}
The inner products evaluate simply to $\braket{g|\pm} = \pm 1/\sqrt{2}$, allowing isolation of the motional operators and simplification using Euler's identity:
\begin{equation}
\begin{split}
    \braket{\hat\sigma_z} &= -\frac{1}{2}\bra{\psi}
        \left[e^{i\theta}\hat{\mathcal{D}}(-\beta)+e^{-i\theta}\hat{\mathcal{D}}(\beta)\right]
    \ket{\psi}\\
    &= -\frac{1}{2}\bra{\psi}
        \left[(\cos\theta+i\sin\theta)\hat{\mathcal{D}}(-\beta)+(\cos\theta-i\sin\theta)\hat{\mathcal{D}}(\beta)\right]
    \ket{\psi}\\
    &= -\frac{1}{2}\bra{\psi}
        \left[\cos\theta(\hat{\mathcal{D}}(\beta)+\hat{\mathcal{D}}(-\beta))-i\sin\theta(\hat{\mathcal{D}}(\beta)-\hat{\mathcal{D}}(-\beta))\right]
    \ket{\psi}\\
    &= -\frac{1}{2}\left[
        \cos\theta\left(\chi(\beta)+\chi^*(\beta)\right)+\sin\theta\left(\chi(\beta)-\chi^*(\beta)\right)
    \right]\\
    &= -\cos\theta\mathfrak{Re}\left[\chi(\beta)\right]-\sin\theta\mathfrak{Im}\left[\chi(\beta)\right],
\end{split}
\end{equation}
\noindent{where} we invoked the definition of the characteristic function $\chi(\beta) = \bra{\psi} \hat{\mathcal{D}}(\beta) \ket{\psi}$ and the identity for complex number $z$ where $\mathfrak{Re}[z] = (z + z^*)/2$ and $\mathfrak{Im}[z] = (z - z^* )/2$, where $z^*$ is the complex conjugate of $z$. Therefore, the choice of single-qubit rotation angle $\theta$ provides  access to a linear combination of the characteristic function's real and imaginary parts. Choosing $\theta=0$ ($\theta = \pi/2$) yields the real (imaginary) part of $\chi(\beta)$. The complex displacement parameter $\beta$ is precisely the argument of the motional state's characteristic function.    
\end{appendices}

\bibliography{sample}

\end{document}